\documentclass[aps,prc,floatfix,showpacs,twocolumn, preprintnumbers, nofootinbib]{revtex4-1}
\usepackage{dcolumn}
\usepackage{bm}
\usepackage{color}
\usepackage{amssymb}
\usepackage{amsmath}
\usepackage{graphicx}
\usepackage{amsfonts}
\usepackage{slashed}
\usepackage{pstricks}
\usepackage{float}
\usepackage[colorlinks = true,
            linkcolor = blue,
            urlcolor  = blue,
            citecolor = blue,
            anchorcolor = blue]{hyperref}
\usepackage{array}
\allowdisplaybreaks
\usepackage{dcolumn}
\usepackage{epsf}
\usepackage{slashed}
\usepackage{tikz}
\usepackage{subfig}
\usepackage{comment}
\graphicspath{{./}{./figures/}}

\usetikzlibrary{shapes, arrows.meta, chains, scopes, positioning, matrix, calc, external}
\usepackage{color}
\definecolor{BrickRed}{rgb}{0.8, 0.25, 0.33}
\definecolor{gray}{rgb}{0.6,0.6,0.6}
\definecolor{darkgreen}{rgb}{0.0, 0.545098, 0.0}
\definecolor{mypink1}{rgb}{0.858, 0.188, 0.478}

\newcommand{\GeV}{~\ensuremath{\text{GeV}}}   
\newcommand{\MeV}{~\ensuremath{\text{MeV}}}  
\newcommand{\fm}{~\ensuremath{\text{fm}}} 

\mathchardef\mhyphen="2D 
\usepackage{scalefnt,xspace}
\newcommand{\myscalefont}[1]{{\scalefont{0.8}#1}}
\newcommand{\achilles}{A\myscalefont{CHILLES}\xspace}

\begin{document}
\preprint{FERMILAB-PUB-22-411-T, MIT-CTP/5428}
\title{\texorpdfstring{\achilles}{Achilles}: A novel event generator for electron- and neutrino-nucleus scattering}
\author{
{Joshua} Isaacson$^{\, {\rm a} }$,
{William} I. Jay$^{\, {\rm d} }$,
{Alessandro} Lovato$^{\, {\rm b,c} }$,
{Pedro} A. N. Machado$^{\, {\rm a} }$,
{Noemi} Rocco$^{\, {\rm a} }$,
}
\affiliation{
$^{\,{\rm a}}$\mbox{Theoretical Physics Department, Fermi National Accelerator Laboratory, P.O. Box 500, Batavia, IL 60510, USA}\\
$^{\,{\rm b}}$\mbox{Physics Division, Argonne National Laboratory, Argonne, Illinois 60439, USA}\\
$^{\,{\rm c}}$\mbox{INFN-TIFPA Trento Institute of Fundamental Physics and Applications, Via Sommarive, 14, 38123 {Trento}, Italy}\\
$^{\,{\rm d}}$\mbox{Center for Theoretical Physics, Massachusetts Institute of Technology, Cambridge, MA 02139, USA}\\
}
\date{\today}

%
\date{\today}
\begin{abstract} 
We present a novel lepton-nucleus event generator: \achilles, {\bf A} {\bf CHI}cago{\bf L}and {\bf L}epton {\bf E}vent {\bf S}imulator. 
The generator factorizes the primary interaction from the propagation of hadrons in the nucleus, which allows for a great deal of modularity, facilitating further improvements and interfaces with existing codes.
We validate our generator against high quality electron-carbon scattering data in the quasielastic regime, including the recent CLAS/e4v reanalysis of existing data.
We find good agreement in both inclusive and exclusive distributions.
By varying the assumptions on the propagation of knocked out nucleons throughout the nucleus, we estimate a component of theoretical uncertainties.
We also propose novel observables that will allow for further testing of lepton-nucleus scattering models.
\achilles\ is readily extendable to generate neutrino-nucleus scattering events.
\end{abstract}
\maketitle

\section{Introduction} \label{sec:intro}

Interactions between leptons with nuclei at beam energies in the 10~MeV$-$10~GeV range are key to properly interpret a multitude of experiments, e.g. those that probe neutrino oscillations~\cite{Itow:2001ee, Ayres:2004js, Acciarri:2015uup, Abe:2018uyc}, electron-nucleus scattering~\cite{Abbott:1997bc, Dutta:2000sn, Mecking:2003zu, Dutta:2003yt, Rohe:2005vc, Burkert:2020akg}, dark matter and dark sectors or even muon-specific new gauge forces~\cite{Battaglieri:2017aum, Kahn:2018cqs, Akesson:2018vlm, Battaglieri:2019nok, Agrawal:2021dbo, Battaglieri:2021rwp}. 
Nevertheless, modeling these interactions with the percent-level precision required by experimental analyses~\cite{Acciarri:2015uup} is a formidable challenge.
These difficulties are primarily due to non-perturbative nuclear dynamics, which play an important role in both the hard interaction vertex and in the propagation of the struck nucleons before they exit the nucleus~\cite{Serber:1947zza,Metropolis:1958sb, Bertini:1963zzc,Cugnon:1980zz,Bertsch:1984gb,Stoecker:1986ci, Bauer:1986zz, Bertsch:1988ik, Botermans:1990qi, Cassing:1990dr, Danielewicz:1991dh, Teis:1996kx, Cugnon:1996xf, Golubeva:1997at, Boudard:2002yn, Hayato:2002sd, Casper:2002sd, Duarte:2007jd, Andreopoulos:2009rq, Iwamoto:2010zzb, Buss:2011mx, Golan:2012wx, Sawada:2012hk, Uozumi:2012fm, Battistoni:2013tra,Battistoni:2015epi,  Mosel:2019vhx, Niewczas:2019fro, Isaacson:2020wlx, Dytman:2021ohr}.

Accurately modeling the hard-scattering cross section between a lepton and a nuclear constituent presents non-trivial difficulties.
The individual couplings between leptons and nucleons are parametrized in terms of single-nucleon form factors, which depend on the momentum transfer associated with the process and involve non-perturbative QCD dynamics.
These form factors can be either calculated from first principles using lattice QCD or fitted to experimental data (see e.g. Refs.~\cite{Bhattacharya:2011ah, Meyer:2016oeg, Davoudi:2020ngi, Borah:2020gte, Meyer:2022mix} and references therein).
Whichever method is used, care must be taken in properly estimating uncertainties, especially in the axial sector where experimental data are scarce. 
In addition, scattering does not occur on a collection of free nucleons (which could be described trivially within a Fermi-gas model).
Instead, the real-world target nucleus is a correlated quantum many-body system; hence, a percent-level theoretical description of scattering off of a bound nucleon must capture many-body correlation effects within a well-defined factorization scheme like the impulse approximation and its generalizations (see e.g. Refs.~\cite{Benhar:2006wy, Martini:2009uj, Martini:2010ex, Amaro:2010sd,Gran:2013kda, Benhar:2015ula}).

The final-state interactions (FSI) that the struck nucleons undergo before exiting the nucleus are also extremely complex phenomena~\cite{Serber:1947zza,Metropolis:1958sb,Bertini:1963zzc,Cugnon:1980zz}, subject to non-perturbative single and many-nucleon effects. 
Examples of the former are nucleon excitations into a $\Delta$ isobar, which quickly decays into a pion-nucleon state.
The latter include correlations induced by realistic two- and three-nucleon forces.
A fully quantum mechanical description of these processes presents an exponentially hard computational problem.
Sophisticated nuclear many-body methods leverage leadership-class computing resources to tackle this real-time nuclear dynamics problem but are limited to inclusive processes and to the non-relativistic regime~\cite{Lovato:2020kba,Sobczyk:2021dwm}.
Exploratory calculations on quantum devices show promise~\cite{Roggero:2018hrn, Roggero:2019myu} for treating fully-exclusive processes.
However, the inclusion of relativistic effects poses non-trivial challenges, and their application to realistic systems seems to remain a distant goal.

Over the years, a number of complementary methods have been developed to capture the leading effects of FSI.
The most sophisticated ones start from the Kadanoff-Baym integro-differential equations for the evolution of the entire nuclear system.
In practice, state-of-the-art transport codes solve truncated versions of these equations, which in turn make a proper estimation of theory uncertainties much harder~\cite{KadanoffBaym:1962, Botermans:1990qi, Cassing:1990dr,Teis:1996kx, Buss:2011mx,Mosel:2019vhx}.
On the other hand, intranuclear cascade (INC) approaches approximately solve the transport equation by evaluating the collision term stochastically.
The main approximation of INC models is that of classical propagation between consecutive quantum-mechanical scatterings. Hence, they are applicable in the regime in which the de Broglie wavelength of the nucleons is much smaller than the range of the interaction, which is in turn smaller than the average distance between nucleons~\cite{Cugnon:2003}. Hence, the applicability of INCs is in principle restricted to nucleons with kinetic energies above $\sim 200$ MeV, although many observables in heavy-ion reactions at less than $100$ MeV are reproduced well in practice~\cite{Dore:2001kn,Uozumi:2012fm}. 

Besides the aforementioned intrinsic limitations, INCs involve a number of additional model-specific prescriptions. For instance, the kinematic variables of a struck nucleon are drawn from a model of the nuclear ground-state; typically the (local or global) Fermi gas or realistic spectral functions. In addition, some assumptions for the propagation in the nuclear medium are made. Examples of the latter are the use of mean free path estimates obtained from the nuclear density or propagating nucleons as if they were ``hard-spheres'' with a radius that is proportional to the square root of the total nucleon-nucleon cross section. Finally, corrections due to the nuclear environment are also typically included by means of average nuclear potentials and imposing Pauli blocking in nucleon-nucleon collisions. 

To assess the reliability of such \emph{effective} descriptions, INCs should be systematically and extensively validated against available experimental data.
Electron-nuclei scattering experiments offer large, high-quality data samples including a broad range of experimental observables, which can be used both to benchmark different INC models and to gauge the reliability of their assumptions. Due to the interplay of all these effects, modeling lepton-nucleus interactions constitutes a remarkable challenge.  Nevertheless, state-of-the-art neutrino event generators have only recently started comparing their predictions to electron scattering data~\cite{Buss:2008sj, Leitner:2008ue, Mosel:2018qmv, Isaacson:2020wlx, Ankowski:2020qbe, Dytman:2021ohr, CLAS:2021neh}. The results of these initial comparisons reveal that existing event generators do not describe electron-nucleus scattering data, and consequently neutrino-nucleus scattering data, to the precision level required by next-generation experiments, such as DUNE~\cite{DUNE:2020lwj, DUNE:2020ypp}.

In this paper we take a first step towards a full-fledged lepton-nucleus event generator: \achilles, \textbf{A} \textbf{CHI}cago\textbf{L}and \textbf{L}epton \textbf{E}vent \textbf{S}imulator.
Three aspects of \achilles'~design are worth highlighting explicitly.
First, as  neutrino physics  enters the precision era, we expect that event generators will need to incorporate many technical improvements and new physical insights.
The need for robust, quickly extensible codebases is therefore acute, presenting a challenge not only in scientific computing but also in software engineering. 
To help solve this inherent difficulty, one of the core design principles of our code is \emph{modularity}. 
We have endeavored to divide the code clearly into individual pieces that describe the different physics processes within lepton-nucleus scattering. 
Examples of the constituent parts include the description of the initial state of the nucleus; the ``hard scattering'' between the lepton and constituents of the nucleus; the intranuclear cascade process; and the nuclear potential. 
The advantage of modularity is obvious: an improvement on a specific part of the code requires minimal effort, since all parts are independent.
Well-implemented modular design thus facilitates future improvements and keeping abreast with new advances in the description of lepton-nucleus scattering. 
We have also tried to provide a user-friendly interface to available tools for physics beyond the Standard Model (BSM), such as the recent lepton-tensor interface developed by some of the current authors~\cite{Isaacson:2021xty}.
The ultimate goal is to enable \achilles~users to proceed seamlessly from writing down a BSM Lagrangian to generating events.

Second, the physical structure of the scattering problem also imposes important constraints. 
Neutrino-nucleus scattering involves both vector and axial form factors (associated with one- and two-body nuclear electroweak current operators), while electron-nucleus scattering is dominated by vector form factors from photon exchange. 
Thus, any neutrino event generator should be able to describe electron-nucleus scattering data as a special case of the more general problem. 
Therefore, the benchmarking foundation for any lepton-nucleus event generator must be extensive validation against inclusive and semi-inclusive electron-nucleus scattering data.
The present paper is an attempt to begin laying this foundation for \achilles.

Third, to leverage the existing analysis tools developed and maintained by the high energy event generators at the LHC, we adopt the HepMC3 output format~\cite{Buckley:2019xhk}. 
This format allows easy interface with analysis tools such as \texttt{Rivet}~\cite{Buckley:2010ar,Bierlich:2019rhm} and eventually \texttt{Nuisance}~\cite{Stowell:2016jfr}, saving valuable research time for users to focus more on physics and less on coding. 
The HepMC3 output format permits arbitrary parameters to be added to an event, allowing additional event information required by neutrino experiments to be included in a simple and straightforward manner.

This paper constitutes the first step towards the full development of \achilles. 
We perform a comparison between our model of electron-nuclei interactions, including the INC model developed in Ref.~\cite{Isaacson:2020wlx}, against the recent reanalysis of electron-carbon scattering data by the CLAS/e4v collaboration~\cite{CLAS:2021neh}. 
Compared to our previous work, and to obtain more realistic results for exclusive observables, we implement a nuclear potential and simulate the propagation of nucleons within this potential.
We focus on the inclusive quasielastic (QE) cross section, which is better understood than other cross section channels at the energies of interest~\cite{Rocco:2019gfb, Ruso:2022qes}. 
We also consider the angular dependent proton yield,
as well as a few other kinematical observables in the QE regime.

The comparisons performed here test the modelling of the cross section; the impact of initial-state nuclear configurations, particularly those obtained via quantum Monte Carlo methods~\cite{Carlson:2014vla} and the roles of the spectral function, in-medium modifications, and Pauli blocking effects.
Taken together, these comparisons provide valuable insight to the physics of intranuclear cascades.
Besides these comparisons, we also propose new observables that may help in further testing models of electron-nuclei and neutrino-nuclei interactions.
All proposed observables can be readily extracted from existing CLAS data.

The paper is organized as follows.
Sec.~\ref{sec:generator} lays out general considerations for lepton-nucleus scattering. Sec.~\ref{sec:inclusive} compares the results of \achilles to experimental data on inclusive electron-nucleus scattering. Sec.~\ref{sec:potential} describes in medium effects from the nuclear potential. Sec.~\ref{sec:QE_exclusive} provides comparisons to exclusive data. Sec.~\ref{sec:new_observables} proposes novel observables to further test the interaction modelling, following by conclusions in Sec.~\ref{sec:conclusions}.

\section{General lepton-nucleus scattering}
\label{sec:generator}
The general expression of the differential cross section for a scattering process involving a target nucleus $A$ and a lepton $\ell$  leading to a given final state reads
\begin{align}
    d\sigma =& \left(\frac{1}{|v_A-v_\ell|}\frac{1}{4E_A^{\rm in} E_\ell^{\rm in}}\right) \big| \mathcal{M} \big|^2 \nonumber\\
    &\times\prod_f\frac{d^3p_f}{(2\pi)^3}(2\pi)^4 \delta^4\Big(k_A+k_\ell - \sum_f p_f\Big)\,.
    \label{eq:xsec_general}
\end{align}
We denote the initial- and final-state momenta by 
\begin{align}\label{eq:notation}
  &k_\ell^\mu=(E_\ell^{\rm in},{\bf k}) \quad\text{incoming}~\ell,\\
  &k_A^\mu=(E_A^{\rm in},{\bf k}_A)\quad\text{incoming}~A,\\
  &p_f^\mu=(E_f,{\bf p}_f)\quad\text{outgoing~particle~}f,\label{eq:notation-last}
\end{align}
where the index $f$ refers to all the possible hadronic and leptonic final state particles. 
The first term in parenthesis is the flux of incoming particles, with $v_{\ell,A}$ being the velocities, the second encodes the matrix element, and the last line is the phase space for the outgoing particles. In the one-boson exchange approximation, the squared amplitude reads
\begin{equation}
    \big| \mathcal{M} \big|^2 = L_{\mu\nu} W^{\mu\nu} \frac{1}{P^2},
\end{equation}
where $P$ is a generic vector boson propagator, while $L$ and $W$ denote the leptonic and hadronic tensors, respectively.
The leptonic tensor is completely determined by the leptonic process (\textit{e.g.} neutrino charged-current interactions). The hadronic tensor on the other hand contains all information on nuclear dynamics and it is expressed as
\begin{equation}\label{eq:had_tens}
    W_{\mu\nu} = \langle \Psi_0 | J^\dagger_\mu (q) | \Psi_f \rangle \langle \Psi_f | J_\nu (q) | \Psi_0 \rangle\, ,
\end{equation}
where $\Psi_0$ and $\Psi_f$ denotes the hadronic initial and final states, respectively.

Carrying out the full calculation of the many-body wave function and its real-time evolution is an exponentially-hard computational problem. To handle it, the reaction process is modeled by separating the primary interaction vertex from the propagation of the struck particles out of the nucleus. Schematically, this division can be expressed by considering the full matrix element squared as:
\begin{align}
    \label{eq:shower}
    |\mathcal{M}(\{k\} \rightarrow \{p\}) |^2 = \big|\sum\hspace*{-3ex}\int_{\;\;p'} &\mathcal{V}(\{ k\} \rightarrow \{ p'\}) \nonumber\\
&\times \mathcal{P}(\{ p'\}\rightarrow \{ p\}) \big|^2,
\end{align}
where the $\{k\}(\{p\})$ is the set of all initial(final) state particle momenta, 
$\mathcal{V}$ represents the primary interaction vertex producing intermediate particles with momenta 
$\{p'\}$, and $\mathcal{P}$ denotes the time evolution of the intermediate states to the final 
states outside the nucleus. Calculating this equation exactly requires retaining full quantum mechanical 
interference between the primary interaction vertex and the subsequent re-interactions.
Traditionally, due to the complexity of solving Eq.~\eqref{eq:shower} exactly,
the calculation factorizes the two-step process as an incoherent product:
\begin{align}
\label{eq:showerfact}
|\mathcal{M}(\{ k\} \rightarrow \{ p\}) |^2 \simeq  \sum\hspace*{-3ex}\int_{\;\;p'} &|\mathcal{V}(\{ k\} \rightarrow \{ p'\})|^2 \nonumber\\
&\times |\mathcal{P}(\{ p'\}\rightarrow \{ p\}) |^2.
\end{align}
This treatment is similar to the approach taken by the collider community when dressing hard-scattering cross 
sections with parton showers (see e.g.~Ref.~\cite{Hoche:2014rga} for a pedagogical discussion).
By construction, this approximation neglects inference between primary interaction vertices that give rise to identical final states while leaving inclusive observables unaffected.
It is expected that these interference effects are subdominant, and a detailed investigation is left to a future work.
In \achilles, the subsequent evolution probability $|\mathcal{P}|^2$ is handled semi-classically using the algorithm developed in Ref.~\cite{Isaacson:2020wlx}.

Eqs.~\eqref{eq:shower} and~\eqref{eq:showerfact} retain full generality for lepton-nucleus scattering, but implementing them in a concrete calculation requires several choices about the relevant degrees of freedom.
First, one must specify the initial-state nuclear constituents $\{k\}$ which participate in the vertex $\mathcal{V}$.
This question is closely related to the choice of a factorization scheme, to which the following section is dedicated.
Second, one must specify the intermediate-state particles which can appear, either from production at the primary interaction vertex $\mathcal{V}$ or in the system's subsequent evolution $\mathcal{P}$.
Briefly stated, the present work restricts to processes in which protons and neutrons are the only active degrees of freedom.
This choice explicitly neglects, e.g., pion production at the primary interaction vertex.
Under this ansatz, the electroweak current of Eq.~\eqref{eq:had_tens} is expanded as a sum of one and two-nucleon operators
\begin{equation}
    J^\mu(q) = \sum_i j^\mu_i(q) + \sum_{i<j} j^\mu_{ij}(q)\, .
    \label{eq:currents}
\end{equation}
Three- and higher-body terms have been found to be small~\cite{Marcucci:2005zc} and are thus neglected here. 
Generalizing these expressions to other mediators, such as scalars, is straightforward. 

\subsection{Factorization scheme}
As mentioned above, this work focuses on a lepton scattering on a nucleus in the quasielastic regime, in which the dominant reaction mechanism is assumed to be single-nucleon knockout: $\ell + A \rightarrow \ell^\prime + X$ where $\ell$ and $\ell^\prime$ denote the initial and final lepton states, $A$ is the target nucleus, $X$ is the hadronic final state, which for example can be composed of a single emitted nucleon and the remnant nucleus.
We can rewrite the cross section of Eq.~\eqref{eq:xsec_general} for a $2\rightarrow 2$ scattering process in a background as
\begin{align}
    d\sigma = & \frac{d^3p_\ell}{(2\pi)^3}\frac{1}{2 E_{\ell}} \frac{d^3 p_X}{(2\pi)^3}\frac{1}{2 E_X} L_{\mu\nu}W^{\mu\nu}\frac{1}{P^2} 
    \label{eq:xec:2to2}
\end{align}
where again $p$ refer to final state momenta, with $X$ denoting the hadronic final state, and
the factor $1/|v_A-v_\ell| =1$ in the lab frame (neglecting the lepton mass).
Note that, for convenience, we absorbed the factors $1/2E_{A (\ell)}^{\rm in}$ in the hadronic (leptonic) tensor, and we are also embedding the delta function in the hadronic tensor. 
For large enough values of the momentum transfer, the virtual boson primarily interacts with individual bound nucleons, so that the hadronic final state can be approximated by the factorized expression
\begin{equation}
    |\Psi_f\rangle = |p \rangle \otimes |\Psi_f^{A-1}\rangle\, ,
\end{equation}
where $|p\rangle$ is a plane wave describing the propagation of the final state nucleon with momentum $|{\bf p}|$, while $|\Psi_f\rangle_{A-1}$ denotes the $(A-1)$-body spectator system, which can be either in a bound or unbound state.
In addition, since we are focusing on the primary interaction vertex, we have dropped the prime in the intermediate variables. 

Retaining the one-body current contribution only in Eq.~\eqref{eq:currents}, the incoherent contribution to the hadronic tensor is given by 
\begin{align}
  W^{\mu\nu}({\bf q}, \omega)&=\!\! \sum_{h \in \{p,n\}} \int  \frac{d^3k_h}{(2\pi)^3}\frac{1}{2E_h^{\rm in}} dE^\prime S_h({\bf k}_h,E^\prime) \nonumber\\
 & \times \langle k| {j_{1b}^\mu}^\dagger|p \rangle \langle p| j_{1b}^\nu| k \rangle (2\pi)^4 \delta^3({\bf k}_h^{\rm}+{\bf q}-{\bf p}_h)\nonumber\\ 
 & \times \delta(\omega-E^\prime+m_N-E_h)\,
 \label{eq:hadronic_tensor}
\end{align}
In the previous equation, ${\bf q}={\bf p}_h-{\bf k}_h$ is the momentum transfer, $k_h=(E_h^{\rm in},{\bf k}_h)$ and $p_h=(E_h,{\bf p}_h)$ denote the energy and momentum of the initial and final nucleons, respectively.
Note that, for brevity, we have suppressed the subscript $h$ in the bras and kets.
The spectral function yields the probability distribution of removing a ``hole'' nucleon $h\in \{p,n\}$ with momentum ${\bf k}_h$ from the target nucleus, leaving the residual $(A-1)$ system with an excitation energy $E^\prime$, and it is defined as~\cite{Benhar:1993ja}
\begin{align}\label{eq:SF}
S_h(&{\bf k}_h, E^\prime) = \\ 
&\sum_{f_{A-1}} |\langle \Psi_0| |k\rangle \otimes |\Psi_f^{A-1}\rangle|^2 \delta(E^\prime+E_0^A-E_f^{A-1}),\nonumber
\end{align}
where the sum runs through the possible final states of the $A-1$  spectator nucleons, which can either be bound or in the continuum.

The spectral function of finite nuclei is generally expressed as a sum of a mean-field and a correlation contribution. 
The first one describes the low momentum and removal-energy region, and is associated with the residual $A-1$ system being in a bound state. 
The correlation contribution includes unbound $A-1$ states for the spectator system, in which at least one of the spectator nucleons is in the continuum, and it provides strength in the high momentum and energy region.
The nuclear spectral function has been evaluated within different semi-phenomenological~\cite{Benhar:1994hw,Ivanov:2018nlm} and ab-initio many-body methods~\cite{Rocco:2018vbf,Barbieri:2019ual}, including quantum Monte Carlo~\cite{Andreoli:2021cxo}. 

The one employed in this work has been obtained within the correlated basis function theory of Ref.~\cite{Benhar:1994hw}. 
The low momentum and energy contribution is determined by adjusting mean-field calculations to reproduce ($e$, $e^\prime$p) scattering measurements. 
The correlation part is derived within the Local Density Approximation by convoluting the correlation component of the spectral function obtained within the correlated basis function theory for isospin-symmetric nuclear matter for a given value of the density.
Additional details regarding the spectral function and, in particular, corrections to the impulse approximation stemming from final-state interactions appear in Sec.~\ref{inclusive_data}.

The spectral function is normalized as
\begin{align}
    \int \frac{d^3 k_h}{(2\pi)^3} dE^\prime S_h({\bf k}_h,E^\prime) =
    \begin{cases}
    Z,\quad\quad\, h={\rm p},\\
    A-Z,\, h={\rm n}.\end{cases}
\end{align}
where $Z$ denotes the number of protons in the nucleus. After applying the factorization ansatz to the hadronic final state, the phase space factor of Eq.~\eqref{eq:xec:2to2} can be  rewritten as
\begin{equation}
    \frac{d^3p_f}{(2\pi)^3}\frac{1}{2E_f} \rightarrow \frac{d^3p_h}{(2\pi)^3}\frac{1}{2E_h} \sum_{f_{A-1}}
\end{equation}
where the discrete sum over the states of the remnant nucleus is embedded in the spectral function as shown in Eq.~\eqref{eq:SF}. 

A complete estimate of the theoretical uncertainty associated with the cross section calculation would require assessing the error in the many-body calculation of the spectral function, the inputs used to describe the interaction vertex (i.e. couplings, form factors), and the factorization of the hadronic final state. Achieving this goal is highly nontrivial and has not been included in this work but future developments are discussed in Sec.~\ref{sec:conclusions}.

\section{Inclusive electron-nucleus scattering}
\label{sec:inclusive}
\subsection{Theoretical preliminaries}
\label{sec:cross_section}
Now we proceed to the concrete calculation of the electron-nucleus cross section, which will be the basis of all comparisons between \achilles\ and electron-carbon scattering data.
We focus first on comparisons between our theoretical predictions and experimental data for the quasielastic inclusive electron-$^{12}$C cross section using the aforementioned factorization scheme and spectral function formalism. 
For those kinematics in which FSI are expected to be negligible, the inclusive cross section provides a benchmarking test for the model of the primary interaction ($|\mathcal{V}|^2$ in Eq.~\ref{eq:showerfact}), since this 
 observable is unaffected by the semi-classical propagation in the nuclear medium, and hence by the INC.

The inclusive double differential cross section for the scattering of an electron 
on an at-rest nucleus via one-photon exchange is written as (see Eqs.~(\ref{eq:notation}-\ref{eq:notation-last}) for notation)
\begin{align}
\Big(\frac{d^2\sigma}{d E_e d\Omega_e}\Big) & =\frac{\alpha^2}{Q^4}{E_{e}^2}L_{\mu\nu}W^{\mu\nu}\, ,
\label{xsec:em1}
\end{align}
where 
$\alpha\simeq1/137$ is the fine structure constant, 
and $\Omega_e$ is the scattering solid angle in the direction specified by ${\bf p}_e$.
The energy and the momentum transfer are denoted by $\omega$  and {\bf q}, respectively, with $Q^2=-q^2={\bf q}^2-\omega^2$.
The lepton tensor is fully determined by the lepton kinematic variables and, neglecting the electron mass, it is given by 
\begin{equation}
L^{\mu \nu}  = \frac{1}{E_e^{\rm in} E_e} (k_e^\mu p_e^\nu + p_e^\mu k_e^\nu - g^{\mu\nu}\, k_e \cdot p_e )\, .
\label{eq:lepton_def}
\end{equation}
The one-body electromagnetic current operator entering Eq.~\eqref{eq:hadronic_tensor} is written as
\begin{align}
j^\mu_{1b}= {\mathcal F}_1 \gamma^\mu+ i \sigma^{\mu\nu}q_\nu \frac{{\mathcal F}_2}{2m_N}\ ,
\label{rel:1b:curr}
\end{align}
where the isoscalar (S) and isovector (V) form factors, ${\mathcal F}_1$ and ${\mathcal F}_2$,
are given by combination of the Dirac and Pauli ones, $F_{1}$ and $F_2$, as
\begin{align}
{\mathcal F}_{1,2}= \frac{1}{2}[F_{1,2}^S+F_{1,2}^V\tau_z]\, ,
\end{align}
$\tau_z$ is the isospin operator, and 
\begin{align}
F_{1,2}^S=F_{1,2}^p + F_{1,2}^n,\ \ \ \ \ \ \ \ F_{1,2}^V=F_{1,2}^p - F_{1,2}^n\, . 
\label{f12:iso}
\end{align}
The Dirac and Pauli form factors can be expressed in terms of the electric and magnetic form factors of the proton and neutron as
\begin{align}
F_1^{p,n}=&\frac{G_E^{p,n}+\tau G_M^{p,n}}{1+\tau},\ \ \ \ \ \ \ \ F_2^{p,n}=\frac{G_M^{p,n}-G_E^{p,n}}{1+\tau}
\end{align}
with $\tau=Q^2/4m_N^2$. 
Therefore, the electromagnetic current can be schematically written as $j^\mu_{1b, \rm EM}=j^\mu_{\gamma,S}+j^\mu_{\gamma,z}$ where the first is the isoscar term and the second is the isovector multiplied by the isospin operators $\tau_z$. 
The above set of equations can be readily extended to the electroweak case and higher multiplicity processes;
an automation for arbitrary leptonic tensors was developed in~\cite{Isaacson:2021xty}. 

The use of a realistic spectral function combined with a factorization scheme has proven to  reproduce a large fraction of the available electron scattering data (see Ref.~\cite{Rocco:2020jlx} and references therein).
Over the past few years, the factorization scheme has been extended to account for two-nucleon currents and pion-production mechanisms~\cite{Benhar:2015ula,RuizSimo:2016rtu,Hernandez:2007qq,Kamano:2013iva,Nakamura:2015rta,Kamano:2016bgm}. 
The focus of the present work is the quasielastic region, and we leave the implementation of additional channels to a future work.

\subsection{Comparison to data}
\label{inclusive_data}
\begin{figure*}[th]
    \centering
    \includegraphics[width=0.48\textwidth]{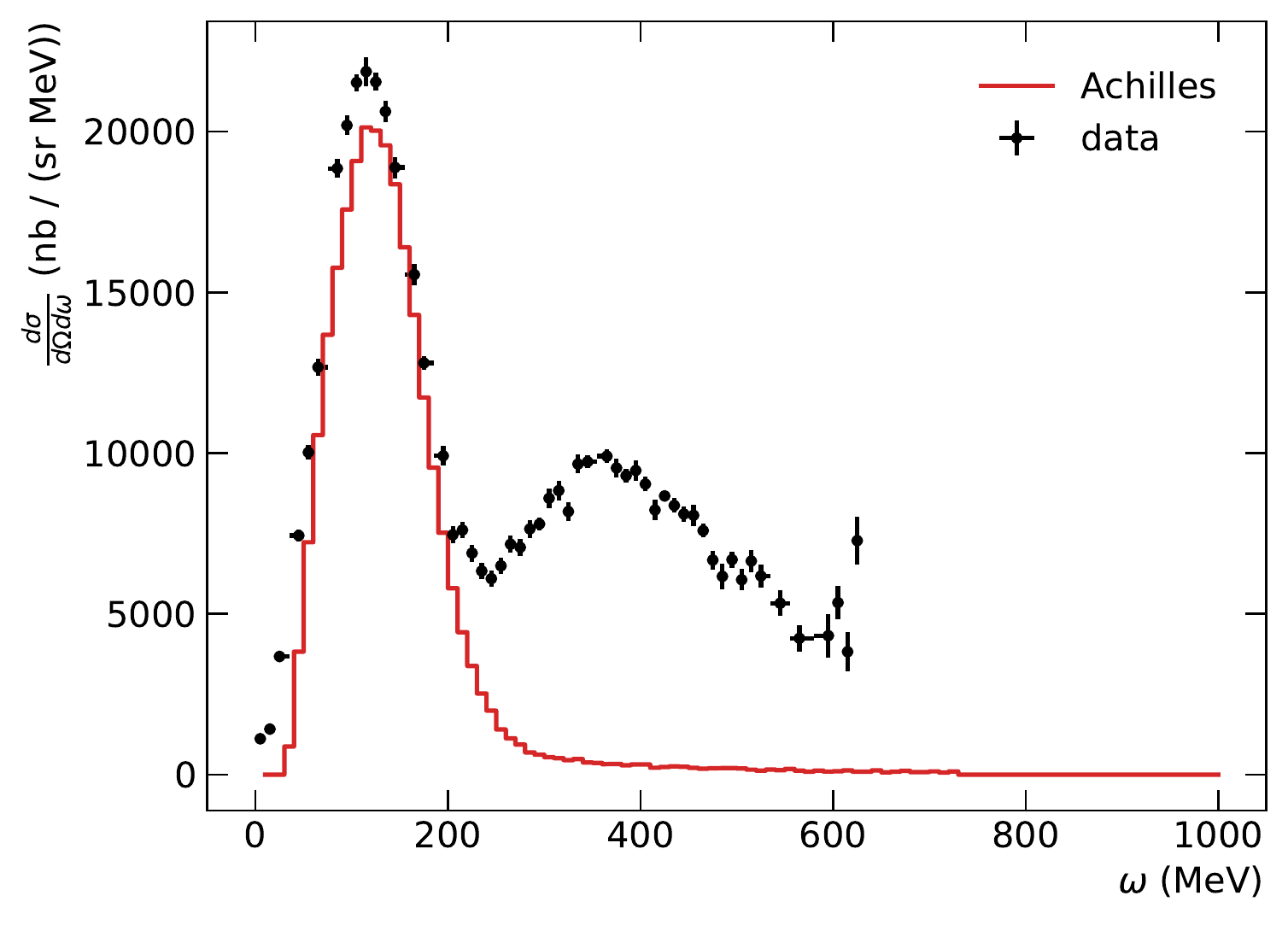}
    \hfill
    \includegraphics[width=0.48\textwidth]{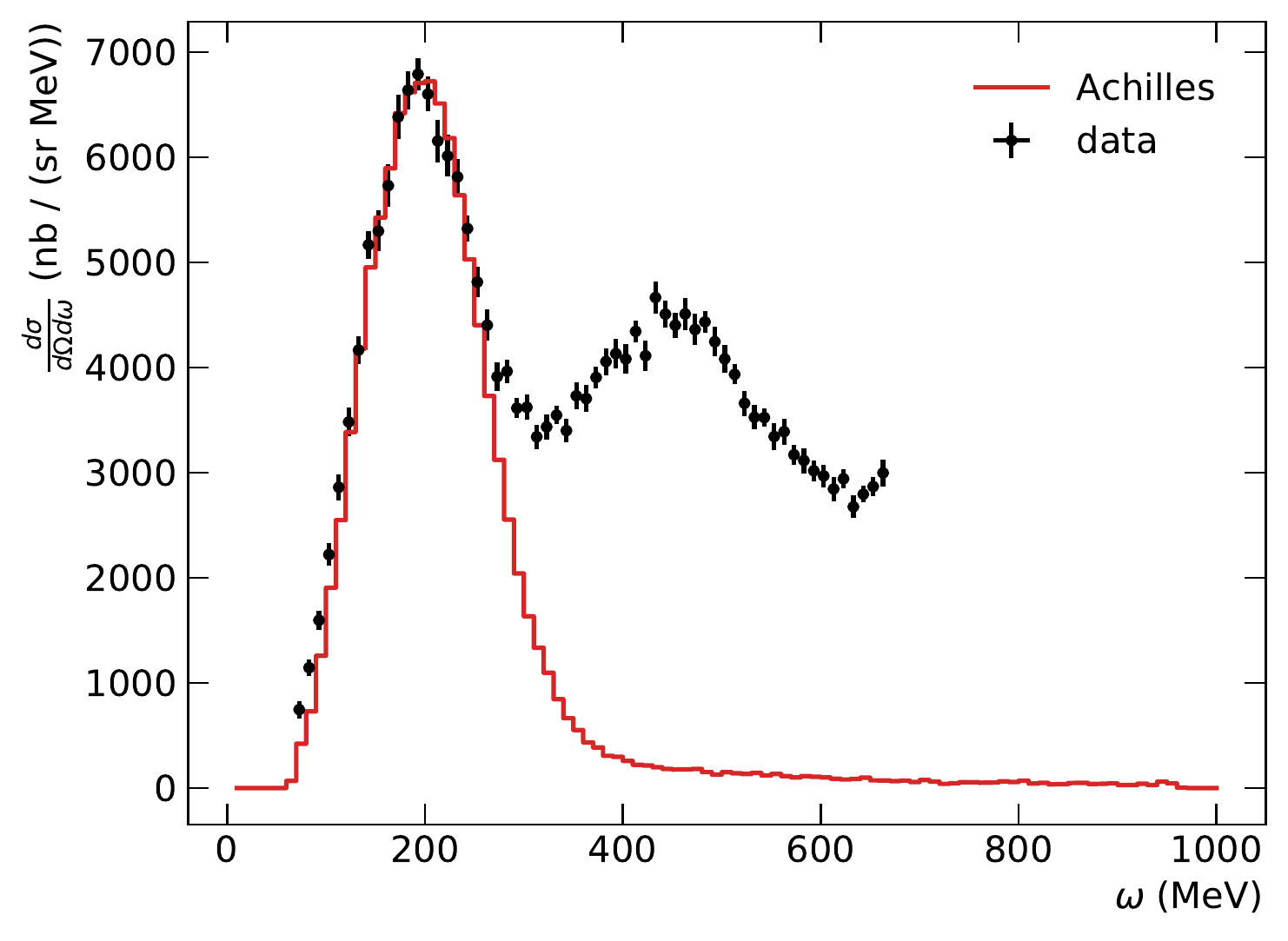}\\
    \includegraphics[width=0.48\textwidth]{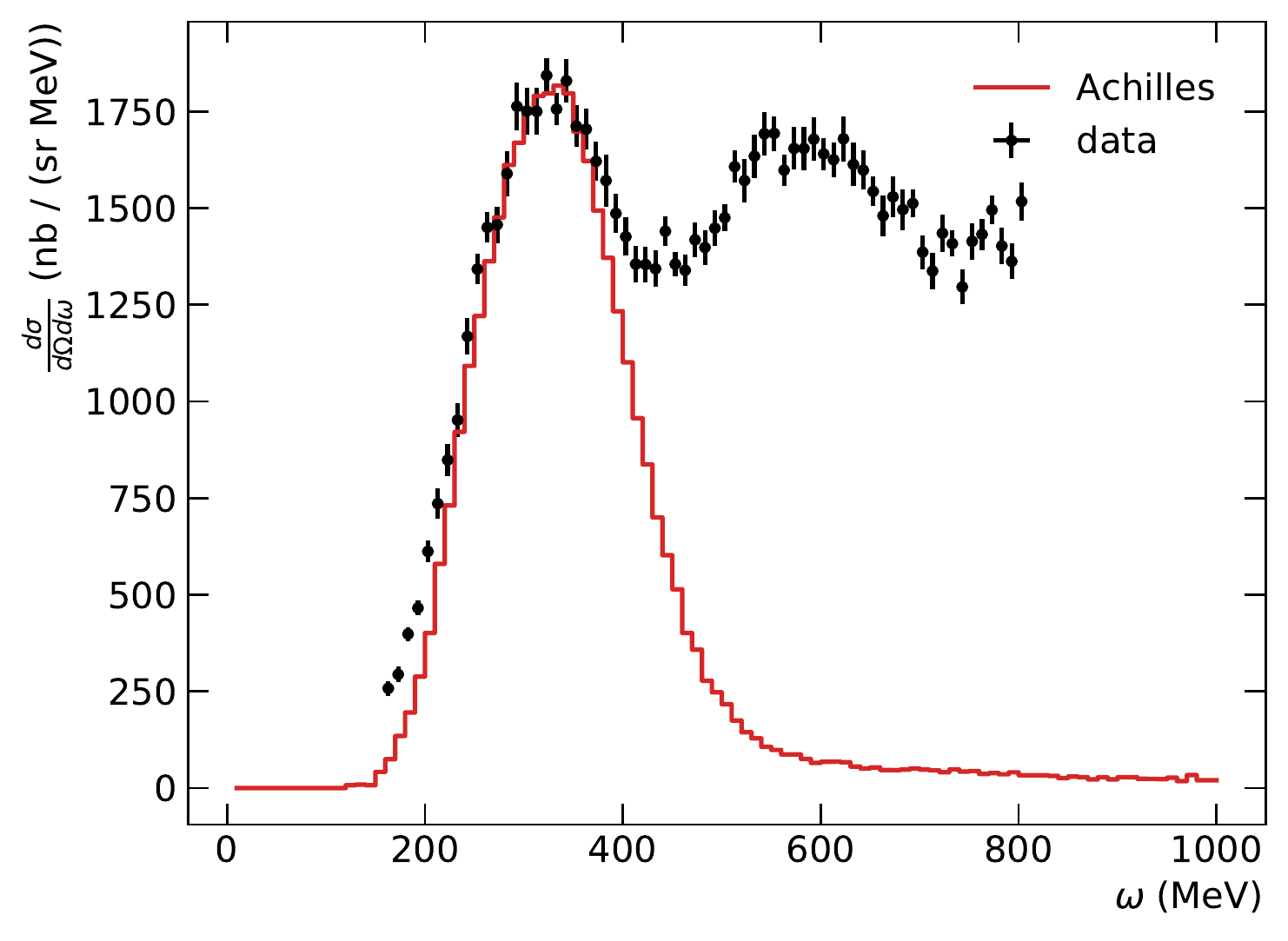}
    \hfill
    \includegraphics[width=0.48\textwidth]{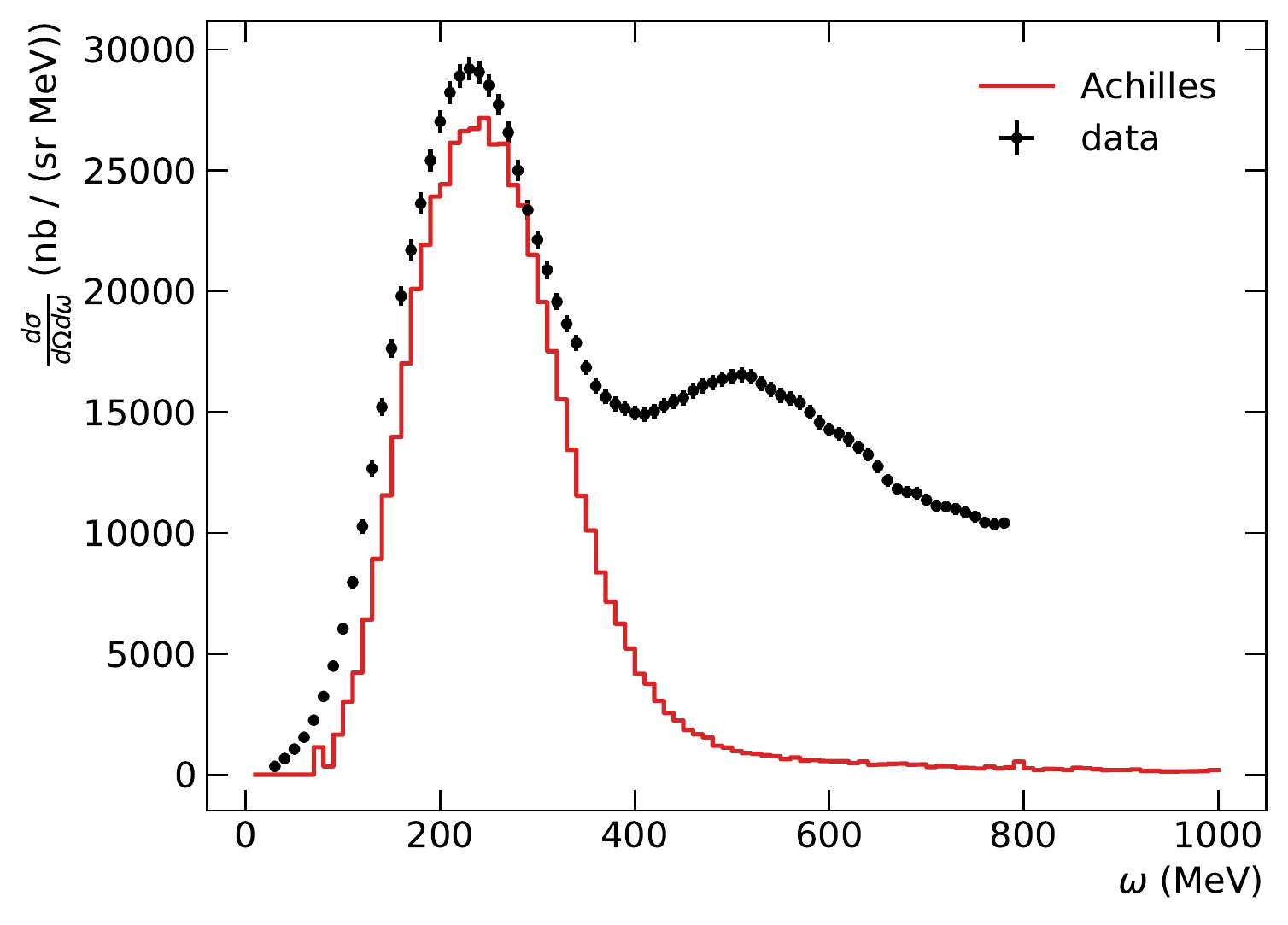}\\
    \caption{Comparison of the \achilles\ event generator to electron-carbon scattering data. Top left: Scattering with an incoming energy of 730 MeV at an angle of 37$^\circ$, data is from~\cite{OConnell:1987kww}. Top right: Scattering with an incoming energy of 961 MeV at an angle of 37.5$^\circ$, data is from~\cite{Sealock:1989nx}. Bottom left: Scattering with an incoming energy of 1300 MeV at an angle of 37.5$^\circ$, data is from~\cite{Sealock:1989nx}. Bottom right: Scattering with an incoming energy of 2500 MeV at an angle of 15$^\circ$, data is from~\cite{Zeller:1973ge}.}
    \label{fig:QE_inclusive}
\end{figure*}

The first comparison between \achilles and data can be found in Fig.~\ref{fig:QE_inclusive} (for technical details of \achilles, see App.~\ref{app:achilles}).
We present the \achilles inclusive $e$-C quasielastic cross section (red histogram) against data as a function of the energy transfer $\omega$. 
Data is taken from several experiments at different incoming electron energy and outgoing electron angle, from top left to bottom right: 730~MeV and 37$^\circ$~\cite{OConnell:1987kww}; 961 MeV and  37.5$^\circ$~\cite{Sealock:1989nx}; 1300 MeVand 37.5$^\circ$~\cite{Sealock:1989nx};  2500 MeV and 15$^\circ$~\cite{Zeller:1973ge}.
In all four cases, the first peak, which is dominated by quasielastic scattering, is quite well described by \achilles.
Note that meson-exchange currents provide additional strength in the dip region between the quasielastic and the resonance peak~\cite{Rocco:2019gfb}. The second peak has large contributions from resonance production, a mechanism which has not yet been implemented in \achilles, and therefore it is not expected to be reproduced by the present version of the code.

Given the large $Q^2$ values of the data displayed in Fig.~\ref{fig:QE_inclusive}, FSI between the struck nucleon and the remnant nucleus are expected to be small and have been neglected in the initial hard interaction. For kinematics in which the factorization scheme is not expected to hold, different approaches have been developed to account for quantum-mechanical effects in FSI in the quasi elastic region. 
To correct the factorization scheme and spectral function results, the real part of an optical nuclear potential~\cite{Cooper:2009zza} is added to the free energy spectrum of the outgoing nucleon and the cross section is convoluted with a \emph{folding function} to account for rescattering effects~\cite{Benhar:2013dq,Benhar:2006wy}. 
In the Relativistic Mean Field approach, FSI between the outgoing nucleon and the residual nucleus are accounted for by solving the associated Dirac equation using the same mean field as used for the bound nucleon~\cite{Gonzalez-Jimenez:2019qhq}.
Including these corrections modifies the inclusive cross section, shifting the quasi elastic peak to lower energy transfers and redistributing the cross section strength to the high-energy-transfer tail~\cite{Ankowski:2014yfa}.

However, the aforementioned approaches do not allow for an accurate treatment of exclusive processes. 
In this regard, INCs are a common tool~\cite{Boudard:2002yn, Hayato:2002sd, Casper:2002sd, Andreopoulos:2009rq, Golan:2012wx,  Battistoni:2013tra, Battistoni:2015epi, Isaacson:2020wlx, Dytman:2021ohr, Ershova:2022jah} for 
modeling the total hadronic state that escapes the nucleus after the hard interaction vertex, as described by Eq.~\eqref{eq:showerfact}, and that could be observed in the detector.
INCs use probabilities to determine if an additional scattering occurs. This probabilistic treatment, at least in current algorithms, neglects interference effects. 
Therefore, by definition INCs leave inclusive observables, such as the differential cross section  displayed in Fig.~\ref{fig:QE_inclusive}, unchanged.

It is important to note that the FSI modeled by folding functions and the FSI modeled by INCs arise from the same physics. 
The major differences between the two approaches are the approximations used to include the imaginary part of the nuclear potential. 
Folding functions account for the effects of FSI including interference effects at the cost of integrating out information on the final state nucleons.
On the other hand, INCs capture the exclusive final state nucleons at the cost of neglecting the interference effects.
Therefore, combining the two calculations in a single code results in effectively double counting the imaginary part of the nuclear potential. 
Implementing interference effects into INCs is beyond the scope of this work.

Having established that our interaction model of quasielastic interactions (i.e., $\mathcal{V}$ in Eq.~\ref{eq:showerfact}) reproduces the experimental data, we next move to comparisons with exclusive observables.

\section{Intranuclear Cascade}\label{sec:INC}

Simulating the propagation of the nucleon involved in the hard scattering out of the nucleus is a vital component of a neutrino event generator.
Intranuclear cascade models are a class of algorithms used to reproduce the imaginary part of the nuclear potential using stochastic Monte Carlo methods.
Traditional techniques do not capture the quantum mechanical components involved in this process. Recently, a new technique for intranuclear cascades has been proposed in Ref.~\cite{Isaacson:2020wlx} to begin capturing these effects.
Figure~\ref{fig:algorithm} shows the programmatic flow of our cascade model. 

In this algorithm, the spatial distribution of neutron and protons are sampled from nuclear configurations obtained from QMC calculations fully retaining correlations effects. 
Their initial momentum is generated according to a local Fermi gas model.
Once the target and the projectile are initialized, the particles are propagated using relativistic kinematics. 
In the simplest approximation, these particles follow straight-lines trajectories, but an option to bend these trajectories using nuclear potentials is discussed in the next section. 
We follow a time-like approach for the propagation; at each step of the propagation $\delta t$ we check if an interaction occurred according to 
the nucleon-nucleon scattering cross section using either a Gaussian or cylindrical probability model depending on the impact parameter.
Originally, the only in-medium effect was taken to be the Pauli principle, and below we discuss updates using the nuclear potential.
We keep two separate lists of ``propagating'' and ``spectators'' particles. 
At the beginning of the event, the projectile is the only propagating particle. 
Afterwards, each particle that has collided with a spectator is promoted to a propagating one, while all the others are still labeled as spectators. 
The particles are propagated until they reach the surface of the nucleus where the nucleon is either recaptured or escapes, based on its energy. 

\tikzset{
  decision/.style={
        diamond,
        draw, thick,
        text width=5.5em,
        text badly centered,
        inner sep=0pt
    },
    block/.style={
        rectangle,
        draw, thick,
        text width=10em,
        text centered,
        rounded corners
    },
    cloud/.style={
        draw,
        ellipse,
        minimum height=2em
    },
    descr/.style={
        fill=white,
        inner sep=2.5pt
    },
    connector/.style={
        -latex,
        font=\scriptsize
    },
    rectangle connector/.style={
        connector,
        to path={(\tikztostart) -- ++(#1,0pt) \tikztonodes |- (\tikztotarget) },
        pos=0.5
    },
    rectangle connector/.default=-2cm,
    straight connector/.style={
        connector,
        to path=--(\tikztotarget) \tikztonodes
    },
    line/.style={>=latex,->,thick},
    evey edge quotes/.style = {auto=right},
}

\begin{figure*}
    \centering
    \tikzsetnextfilename{algorithm}
    \begin{tikzpicture}
        \matrix (m) [matrix of nodes, column sep=1cm, row sep=6mm, align=center, nodes={anchor=center}, nodes in empty cells]    
        {
            |[block]| {Start} & &\\
            |[block]| {Generate nuclear configuration} & &\\
            |[block]| {Handle primary interaction} & &\\
            |[block]| {Evolve propagating nucleon(s) for time $\delta t$} & &\\
            |[decision]| {Possible interaction?} & |[decision]| {Outside nucleus?} & |[block]| {Remove from propagating particle list}\\
            |[block]| {Generate outgoing momenta} &  & \\
            |[decision]| {Pauli Blocked?} & |[block]|{Restore original outgoing momenta} & & \\
            |[block]| {Update outgoing momenta of particles} & &|[block]| {Add particle to list and update formation zone}& \\
        };
        \path[line] (m-1-1) edge (m-2-1);
        \path[line] (m-2-1) edge (m-3-1);
        \path[line] (m-3-1) edge (m-4-1);
        \path[line] (m-4-1) edge (m-5-1);
        \path[line] (m-5-1) edge node[midway, above, text=black]{No} (m-5-2);
        \path[line] (m-5-1) edge node[midway, right, text=black]{Yes} (m-6-1);
        \path[line] (m-5-2) edge node[midway, above, text=black]{Yes} (m-5-3);
        \draw[line] (m-5-2) |- node[midway, right, text=black]{No} (m-4-1);
        \path[line] (m-6-1) edge (m-7-1);
        \path[line] (m-7-1) edge node[midway, above, text=black]{Yes} (m-7-2);
        \path[line] (m-7-2) edge (m-5-2);
        \path[line] (m-7-1) edge node[midway, right, text=black]{No} (m-8-1);
        \path[line] (m-8-1) edge (m-8-3);
        \draw[line] (m-8-3) -- (m-7-3.center) -- (m-6-3.center) -- (m-6-2.center) -- (m-5-2);
    \end{tikzpicture}
    \caption{
        The proposed algorithm for the INC model. This figure is reproduced from Ref.~\cite{Isaacson:2020wlx}.
        \label{fig:algorithm}
    }
\end{figure*}
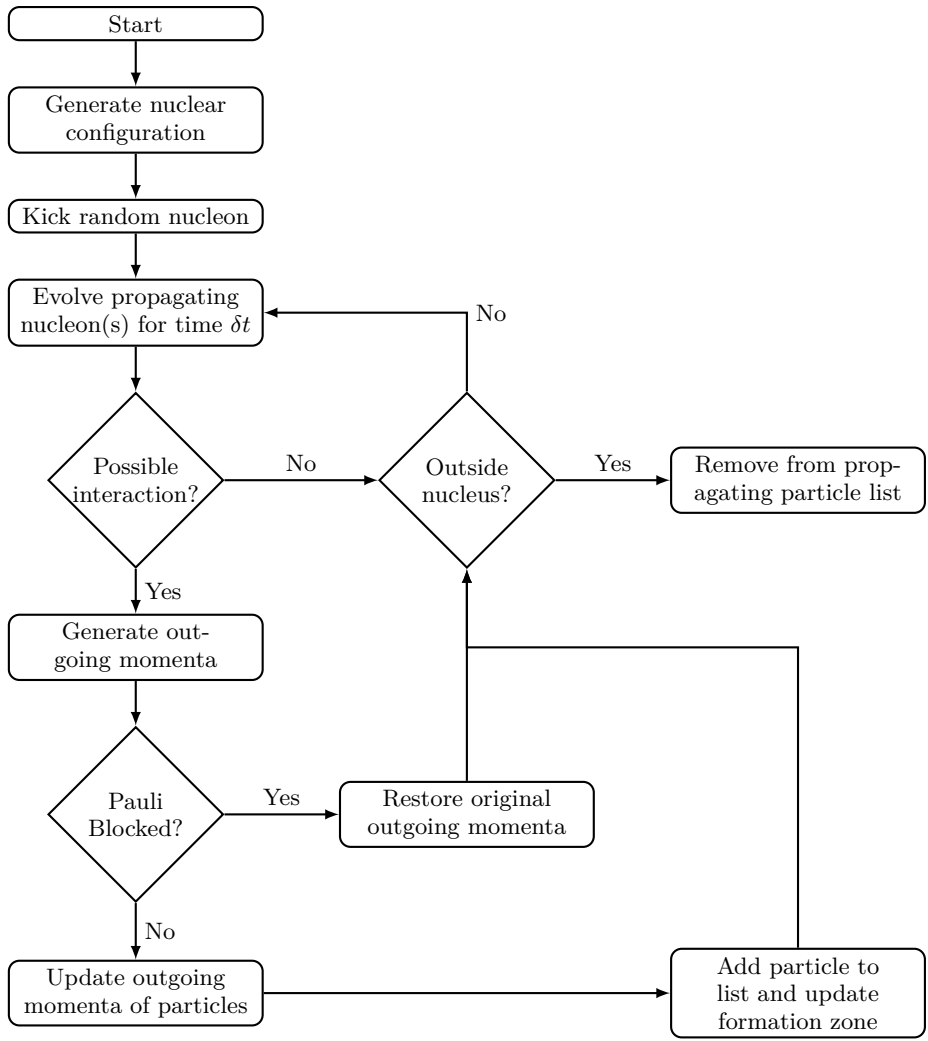

\subsection{Nuclear potential}\label{sec:potential}
In electron-nucleus and neutrino-nucleus scatterings, inclusive quantities may be well described without detailed modelling of what happens when nucleons are propagating out of the nucleus.
The description of exclusive quantities is more demanding.
While nucleon-nucleon interactions are possibly the most important effects to include in an INC model,
the presence of a mean-field nuclear potential may trap struck nucleons or deflect their trajectory, effectively changing the number, momentum and direction of outgoing particles.
To account for this effect, we have implemented two different options as a background potential, which depends on both the position $r$ and momentum $p$ of the propagating nucleon.
Note that, in our approaches, only the real part of the potential is included, since the imaginary part is captured by the hard scattering in the intranuclear cascade. A similar approach of including the potential into cascades was studied in Ref.~\cite{Nikolakopoulos:2022qkq}.
The first potential considered is a non-relativistic potential defined by a three-parameter fit to single-particle energy of infinite nuclear matter~\cite{Wiringa:1988jt}, which is consistent with the variational ground-state calculations of Wiringa, Fiks, and Fabrocini (WFF)~\cite{Wiringa:1988tp}. Its functional form is given as
\begin{equation} 
U(p',r) = \alpha[\rho(r)] + \frac{\beta[\rho(r)]}{1+\left(p'/\Lambda[\rho(r)]\right)^2},
\end{equation}
where $p'$ is the modulus of the three momentum of the propagating nucleon, while $\alpha, \beta$, and $\Lambda$ are fit to reproduce the single-particle energy of nuclear matter as obtained from the Urbana $v_{14}$ + TNI Hamiltonian, and $\rho(r)$ is the local nuclear density at radius $r$. The values of the aforementioned variables are 
\begin{align}
    \alpha(\rho) &= 15.52 (\rho/\rho_0) + 24.93 (\rho/\rho_0)^2 \MeV, \\
    \beta(\rho) &= -116 (\rho/\rho_0) \MeV, \\
    \Lambda(\rho) &= 3.29 - 0.373 (\rho/\rho_0) \fm^{-1},
\end{align}
where $\rho_0=0.16 \fm^{-3}$ is the saturation density of nuclear matter.

The other potential we adopted is based on the work of Ref.~\cite{Cooper:2009zza} where proton-nucleus elastic  and reaction cross section data are fitted
to determine global proton-nucleus optical potentials for energies  between 20 and 1040 MeV for several nuclear targets, including carbon. 
The fitting can be done with potentials in a Dirac equation or Schroedinger equation. 
For the former case, the Dirac equation  was used in the form
\begin{equation}
    \{{\bm \alpha}\cdot{\bf p}'+\beta [m+U_s(r,E')]+U_0(r,E')+V_c(r)]\}\psi({\bf r}) = E' \psi({\bf r}) \nonumber
\end{equation}
where $V_c(r)$ denotes the Coulomb potential at a given nuclear radius $r$, which is either computed from Woods-Saxon-like charge distribution~\cite{Clark:2006rj} or taken from data when they are available, and $E'$ is the energy of the propagating nucleon. 
The quantities determined by the fitting procedure are $U_s(r,E')$ and $U_0(r,E')$, the scalar and vector optical potentials, respectively; they include a real and an imaginary part. To obtain an effective optical potential for the Schroedinger equation, it is helpful to write down a standard reduction of the Dirac equation to second order form. 
The equation for the upper two components is 
\begin{align}
     \{p^{\prime2} +& 2 E'(U_{\rm eff}(r,E') + U_{\rm SO}(r,E')\ { \bf L} \cdot {\bf S})\}\psi_u({\bf r}) \nonumber\\
    & = [(E'-V_c(r))^2-m^2]\psi_u({\bf r})\, ,
\end{align}
where ${ \bf S}$ and ${ \bf L}$ are the total spin and angular momentum of the nucleus, respectively. 
We can identify $U_{\rm eff}$ and $U_{\rm SO}$ as effective Schroedinger-equation central and spin-orbit potentials that can be constructed from $U_s(r,E')$ and $U_0(r,E')$. 
Note that the Schroedinger-equation central potential also includes the Darwin term accounting for relativistic corrections, and its effect is more pronounced in the nuclear interior~\cite{Arnold:1981dt}. The spin-orbit term is significantly smaller than the central one, and for this reason it has been neglected in the present work. In the remainder of this paper, we will denote the potential obtained from Ref.~\cite{Cooper:2009zza} retaining only the central contribution as the Schroedinger potential. 
 
There are two ways that the potential plays a role within the cascade algorithm.
Firstly, the potential modifies the hard interactions that occur between nucleons, often referred to as in-medium modifications in the literature.
In this work, we only consider the non-relativistic in-medium corrections as implemented in Ref.~\cite{PhysRevC.45.791}.
To account for in-medium corrections due to the nuclear potential, we modify the differential cross section using
\begin{align}
    \frac{d\sigma'}{d\Omega} &= \frac{\lvert\mathbf{p}'_1-\mathbf{p}'_2\rvert}{m}
    \left\lvert \frac{\mathbf{p}'_1}{m^*(p'_1,\rho)}-\frac{\mathbf{p}'_2}{m^*(p'_2,\rho)}\right\rvert^{-1} \nonumber \\
    &\times \frac{m^*\left(\sqrt{(p_3^{\prime2}+p_4^{\prime2})/2},\rho\right)}{m}\frac{d\sigma}{d\Omega},
\end{align}
where $p'_1, p'_2$ are the momenta of the incoming propagating nucleons, and
$p'_3, p'_4$ are the momenta of the outgoing propagating nucleons. The effective nucleon mass $m^*$ is given as
\begin{equation}
    m^*(p',\rho) = p' \left(\frac{p'}{m} + \frac{d U(p', \rho)}{dp'}\right)^{-1}.
\end{equation}
This in-medium correction approximates the in-medium matrix element to be the same as the free matrix element, and that $U(p'_1, \rho) + U(p'_2, \rho) \approx U(\sqrt{(p_1^{\prime2}+p_2^{\prime2})/2}, \rho)$.
We leave the expansion to the relativistic case to a future work.
Note that we assume the potential to remain the same regardless of INC dynamics.
While this is certainly an approximation that will fail when the nucleus suffers a ``hard'' breakdown, it should be reasonable when the number of exiting nucleons is much lower than the number of a nucleons in the nucleus.

We also consider the long-distance effect of a background potential on the nucleon as it
propagates through the nucleus.
We simulate a particle propagating  by classical Hamiltonian evolution of the system. The equations of motion can be written as
\begin{align}
    \frac{{\rm d}\mathbf{p}}{{\rm d}t} &= -\frac{\partial H}{\partial \mathbf{q}} =  \left(\frac{(U_s + m) }{\sqrt{\mathbf{p}^2 + (U_s + m)^2}}\frac{\partial U_s}{\partial |\mathbf{q}|} + \frac{\partial U_0}{\partial |\mathbf{q}|}\right) \hat{\mathbf{q}}, \nonumber \\
    \frac{{\rm d}\mathbf{q}}{{\rm d}t} &= \frac{\partial H}{\partial \mathbf{p}} =\left(\frac{(U_s + m)\frac{\partial U_s}{\partial |\mathbf{p}|} + |\mathbf{p}|}{\sqrt{\mathbf{p}^2 + (U_s + m)^2}} + \frac{\partial U_0}{\partial |\mathbf{p}|}\right)\hat{\mathbf{p}}\, .
\end{align}
The equations above are clearly a set of coupled differential equations. 
In order to maintain conservation of energy, a symplectic integrator is used for the evolution. 
Since these differential equations are coupled, traditional symplectic integrators will not work. 
It was shown in Ref.~\cite{PhysRevE.94.043303} that it is possible to use symplectic integrators by working with an augmented Hamiltonian in an extended phase space.
App.~\ref{app:symplectic_int} provides technical details.

\section{Comparison with exclusive observables}\label{sec:QE_exclusive}

We now proceed to the analysis of exclusive observables in electron-carbon scattering.
Exclusive quantities are particularly relevant for neutrino experiments, especially those based on the liquid argon time projection chamber (LArTPC) technology such as the SBN detectors~\cite{MicroBooNE:2015bmn} or the future DUNE experiment~\cite{Acciarri:2015uup}.
LArTPCs are able to identify and reconstruct tracks of all charged particles in a neutrino scattering event in exceptional detail.
This capability allows these detectors to reject backgrounds and optimize searches more efficiently.
If we take as an example the recent MicroBooNE search for single photons~\cite{MicroBooNE:2021zai} as an explanation of the MiniBooNE low energy excess~\cite{MiniBooNE:2020pnu}, we can appreciate the importance of exclusive quantities: the one-photon-zero-proton sample has a background rate 7 times higher than the one-photon-one-proton sample, and this can largely be attributed to the inability to reconstruct the $\Delta \to N\gamma$ invariant mass in the absence of a proton track. 
Several other examples can be made, but the point is that describing correctly exclusive observables will be crucial in current and future neutrino experiments. 

The CLAS and e4v collaborations have recently reported a study of energy reconstruction in electron-nucleus scattering data, using methods employed in neutrino experiments~\cite{CLAS:2021neh}.
The collaborations analyzed electron scattering data taken with CLAS at JLab for three different beam energies: $1.159$, $2.257$, and $4.453\GeV$. 
The detection thresholds for hadrons were similar to thresholds at current and future neutrino experiments.
The analysis focused on the reconstruction of several exclusive and differential quantities, such as incoming electron energy reconstruction for 0$\pi$ events, calorimetric reconstructed energies for 1p0$\pi$ events, transverse variables, proton multiplicity, and so on.

In what follows, we describe the comparison between our generator and CLAS data, for all available observables, focusing on quasielastic electron-carbon scattering.
The CLAS/e4v collaborations have reweighted their data by a factor $Q^4/{\rm GeV}^4$, where $Q^2=-q^2$ is the four-momentum transfer which can be obtained with final and initial electron kinematics. 
This was done to have a better comparison with neutrino events: at these energies, while electron-nucleus scattering is dominated by photon exchange, neutrino-nucleus scattering can be very well approximated by a four-fermion interaction.
Here, we do the same in an event by event basis.

We adopt the same CLAS acceptances and mimic the energy resolution as described in Ref.~\cite{CLAS:2021neh}, after corrections for undetected particles.
The electron and proton energies are smeared by $1.5(0.5)\%$ and $3(1)\%$ for the 1.159 (2.257 and 4.453) GeV beams, respectively.
Protons were detected with momentum $p_p>300\MeV$ and angle with respect to the beam direction $12^\circ<\theta_p$.
Electrons were detected with energy $E_e>0.4,\,0.55,$ and $1.1\GeV$ for $E_{\rm beam}=1.159,\,2.257$ and $4.453\GeV$, as well as angles with respect to the beam direction 
\begin{equation}\label{eq:angle_cut}
  \theta_e^i > \theta_0^i+\frac{\theta_1^i}{p_e~[{\rm GeV}]},
\end{equation} 
where $p_e$ is the electron momentum, $i=1,2,3$ refers to the three beam energies in increasing order, $\theta_0^i=17^\circ, \,16^\circ, \,13.5^\circ$ and $\theta_1^i=7^\circ, \,10.5^\circ, \,15^\circ$.
Since we do not simulate production and propagation of pions, we do not list their acceptances here.

\begin{figure}[t]
    \includegraphics[width=0.4\textwidth]{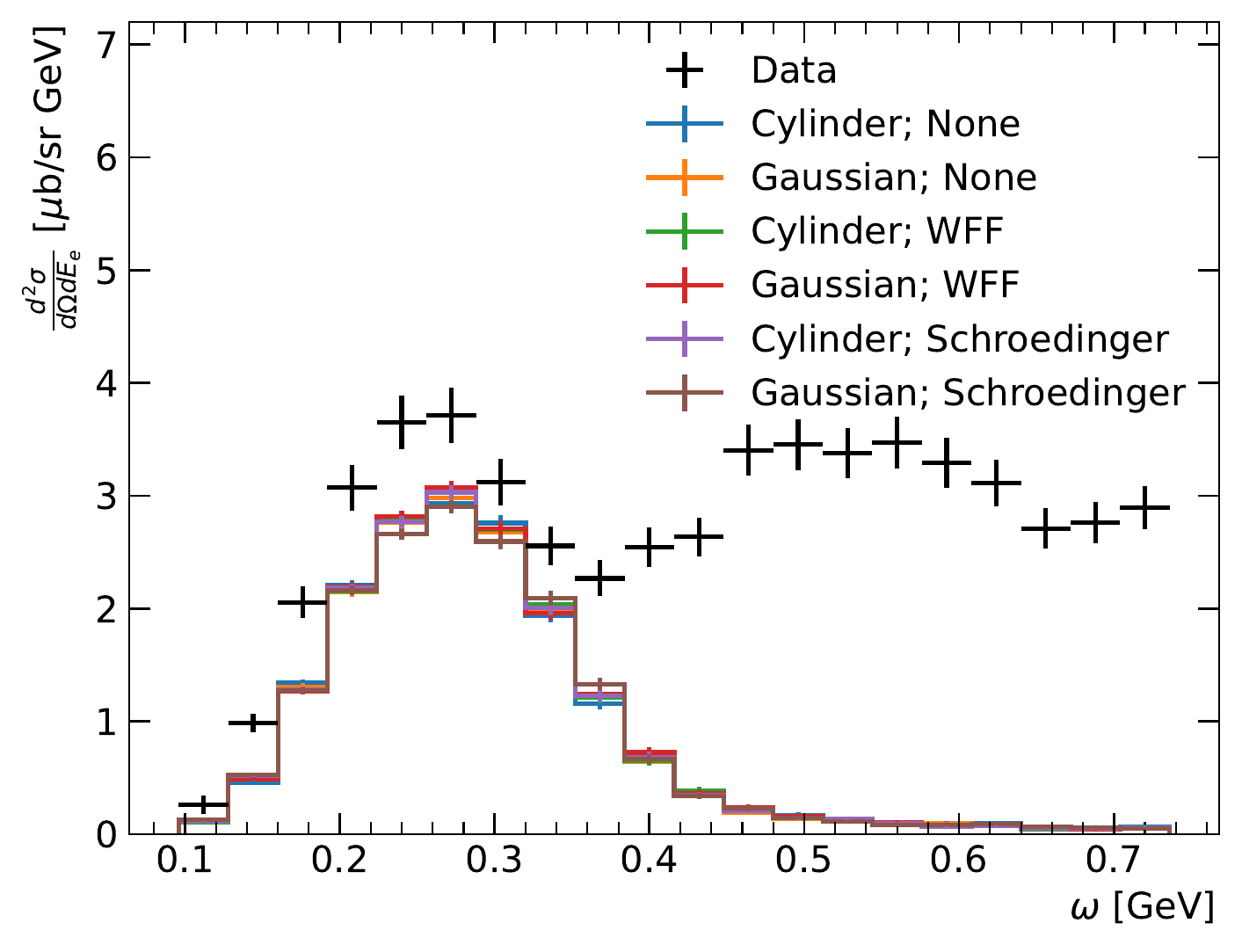}
    \caption{Comparison of the inclusive cross-section for an electron beam of 1159 MeV with outgoing angle of 37.5${}^\circ$. Data is taken from Ref.~\cite{CLAS:2021neh}.}
    \label{fig:e4nu:omega_1159}
\end{figure}

We start with the double differential cross section $d^2\sigma/d\Omega_e dE_e$, where $\Omega_e$ is the solid angle and $E_e$ is the outgoing electron energy, as a function of the energy transfer $\omega\equiv E_e^{\rm in}-E_e$, for fixed outgoing electron angle of $37.5^\circ$ with respect to the beam axis and beam energy $E_e^{\rm in}=1.159$~GeV, see Fig.~\ref{fig:e4nu:omega_1159}.
Hereafter we present \achilles results 
for several different variations on the implementation of the INC, namely, the nucleon-nucleon interaction model (Cylinder vs. Gaussian, see Ref.~\cite{Isaacson:2020wlx}) and the real part of the nuclear potential (WFF, Schroedinger or none). Here ``none,'' is used as a baseline prediction for the model described in Ref.~\cite{Isaacson:2020wlx}.
Different treatments will be color coded and indicated by an inset in all figures.
The spread among the lines can be interpreted as one of the theoretical uncertainties on the lepton-nucleus interaction modeling.
Inclusive observables, such as those displayed in Fig.~\ref{fig:e4nu:omega_1159} are not affected by the semi-classical intranuclear cascades. Therefore, the different lines lie on top of each other.

At low energy transfer, quasielastic scattering dominates the cross section.
In this region, particularly for $0.1<\omega<0.4$~GeV, our generator describes the data fairly well, except for the small-energy region where the theory underestimates the data.
The agreement with data would be improved by the interference effects neglected in intranuclear cascades (as discussed at the end of Sec.~\ref{inclusive_data}), yielding an enhancement of the strength at low $\omega$.
However, a naive combination of our intranuclear cascade and a folding function would result in a double-counting for exclusive observables, as discussed above.
For this reason it has not been included in our calculation.

The missing strength in the quasielastic region and towards higher energy transfers is largely ascribed to meson exchange, resonance production and deep inelastic scattering contributions currently neglected in our analysis~\cite{Rocco:2019gfb}.
As one goes beyond this region towards higher energy transfers, the quasielastic contribution shrinks and one expects other components of the cross section to be more relevant, which explains the discrepancy between the data and our generator.
Overall, this level of agreement is an encouraging result.

\begin{figure}[t]
    \includegraphics[width=0.4\textwidth]{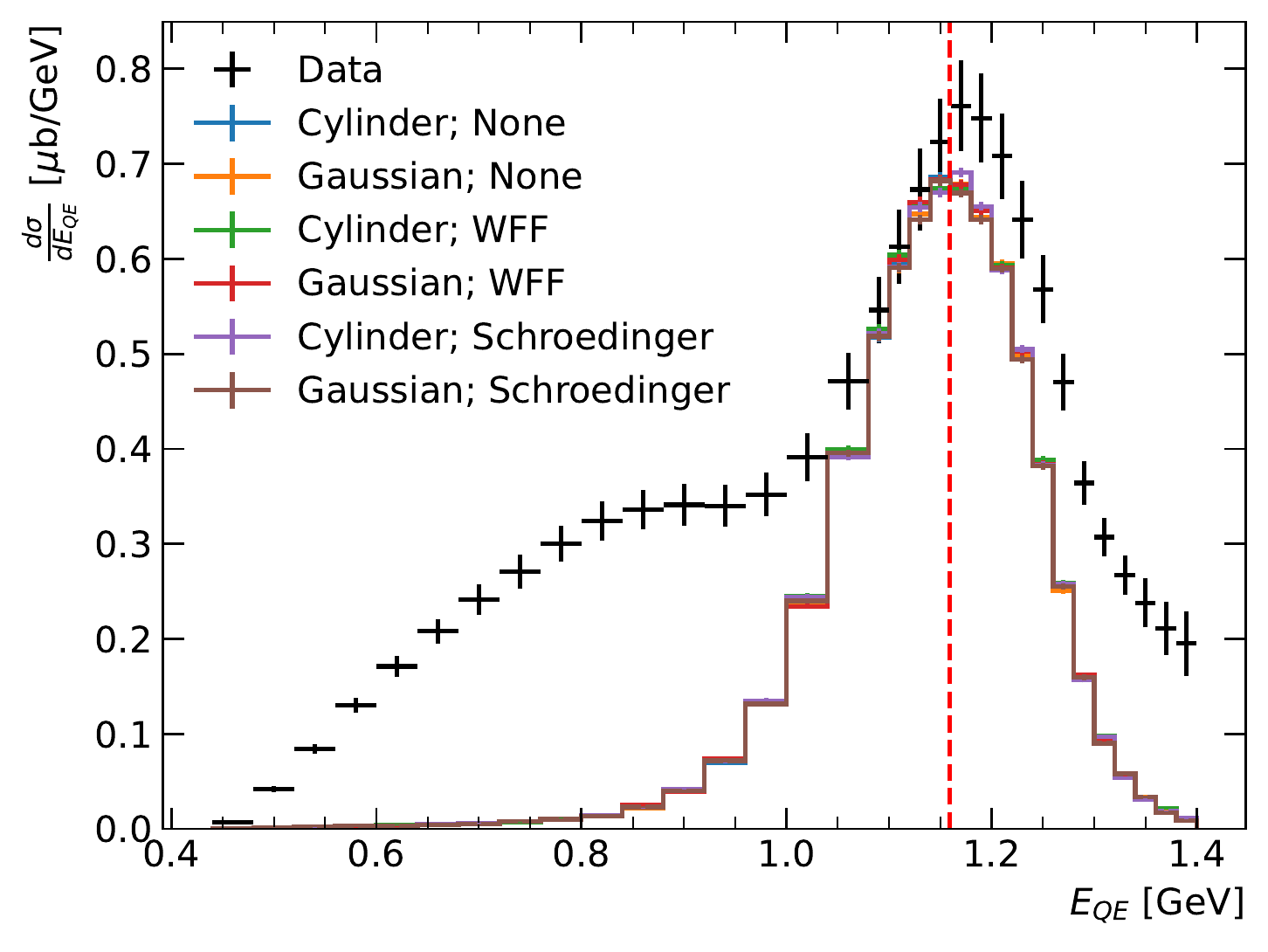}
    \caption{Comparison of the quasielastic energy reconstructed for an electron beam of 1159 MeV.
             Data is taken from Ref.~\cite{CLAS:2021neh}.
             The definition of $E_{QE}$ can be found in Eq.~\ref{eq:eqe}.
             The red dashed vertical line marks the true beam energy.}
    \label{fig:e4nu:e_qe_1159}
\end{figure}

Another comparison we make is on the lepton energy reconstruction assuming quasielastic scattering.
The quasielastic energy reconstruction is done based off the methodology used by water Cherenkov detectors, such as MiniBooNE and T2K.
In this case, only charged leptons and pions are measured. Assuming that the neutrino scatters quasielastically
from a stationary nucleon within a nucleus, its incoming energy can be reconstructed as
\begin{equation}\label{eq:eqe}
    E_{\rm QE} = \frac{2m_N\epsilon+2m_N E_\ell - m_\ell^2}{2\left(m_N-E_\ell+p_\ell \cos\theta_\ell\right)},
\end{equation}
where $m_N$ is the mass of the nucleon, $\epsilon$ is the average nucleon separation energy (we use 21~MeV for carbon),
$E_\ell (p_\ell)$ is the energy (momentum) of the outgoing lepton,
and $\theta_\ell$ is the angle of the outgoing lepton with respect to the beam axis. 
The different scheme choices discussed in this paper are compared to the measured $E_{\rm QE}$
distribution for a 1.159 GeV electron beam on carbon from the CLAS data~\cite{CLAS:2021neh} in
Fig.~\ref{fig:e4nu:e_qe_1159}. 

Here the peak around the beam energy is dominated by the quasielastic 
contribution, while the tail towards lower values of $E_{\rm QE}$ is dominated by meson exchange currents and resonance production.
Therefore, we only expect our results to approximately reproduce the peak, which is what is shown. The agreement with the data for larger values of $E_{\rm QE}$ is likely to be improved by the interference effects neglected by intranuclear cascades. However, a more detailed analysis of the discrepancy will be carried out in the future when meson exchange currents are included in \achilles. Analogously to Fig.~\ref{fig:e4nu:omega_1159}, this distribution has no information about the outgoing protons contained within it.
Therefore, we expect that the prediction should be insensitive to the cascade parameters, as can be seen in the small spread of the colored lines.

\begin{figure}[t]
    \includegraphics[width=0.4\textwidth]{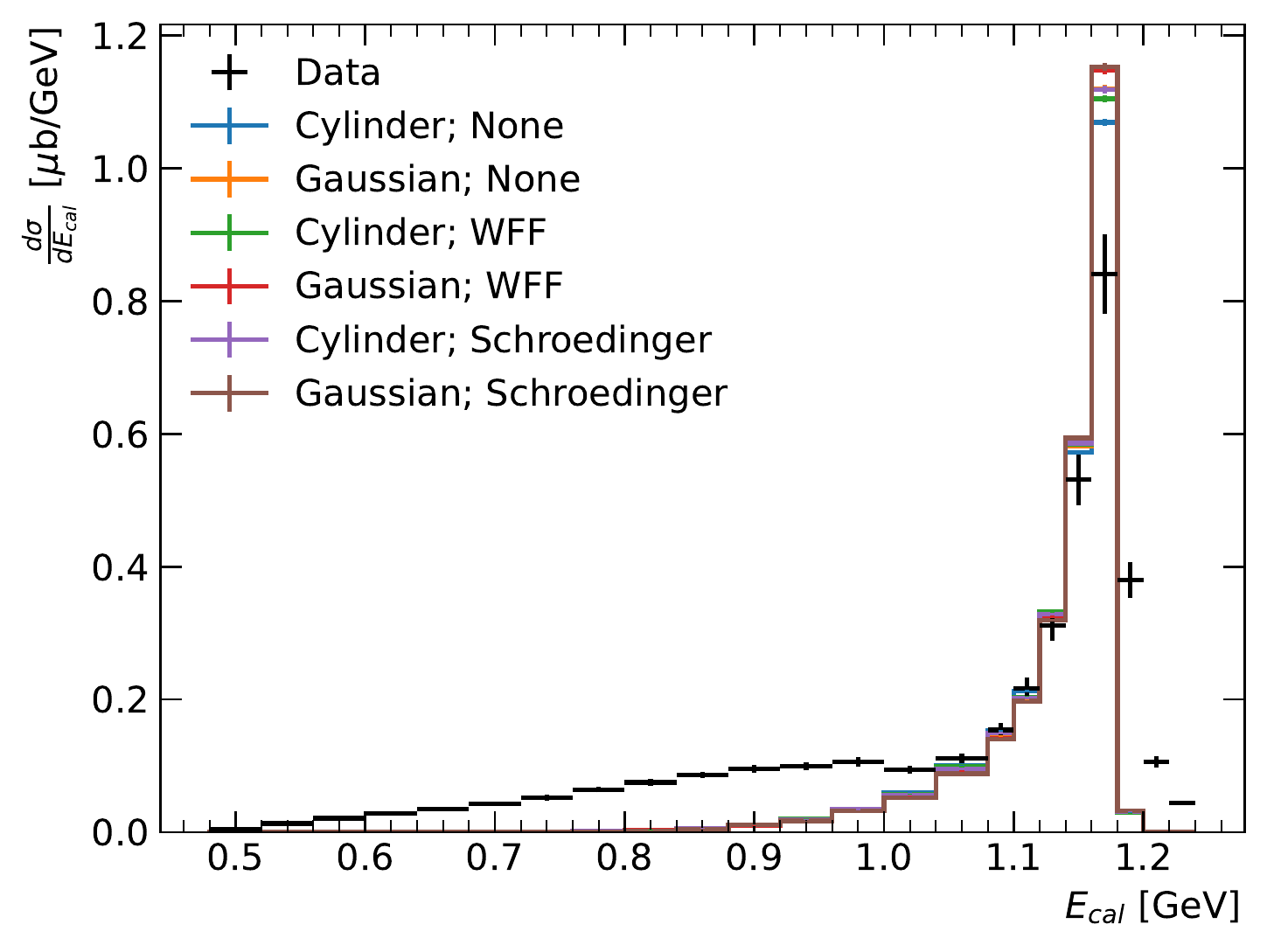}
    \includegraphics[width=0.4\textwidth]{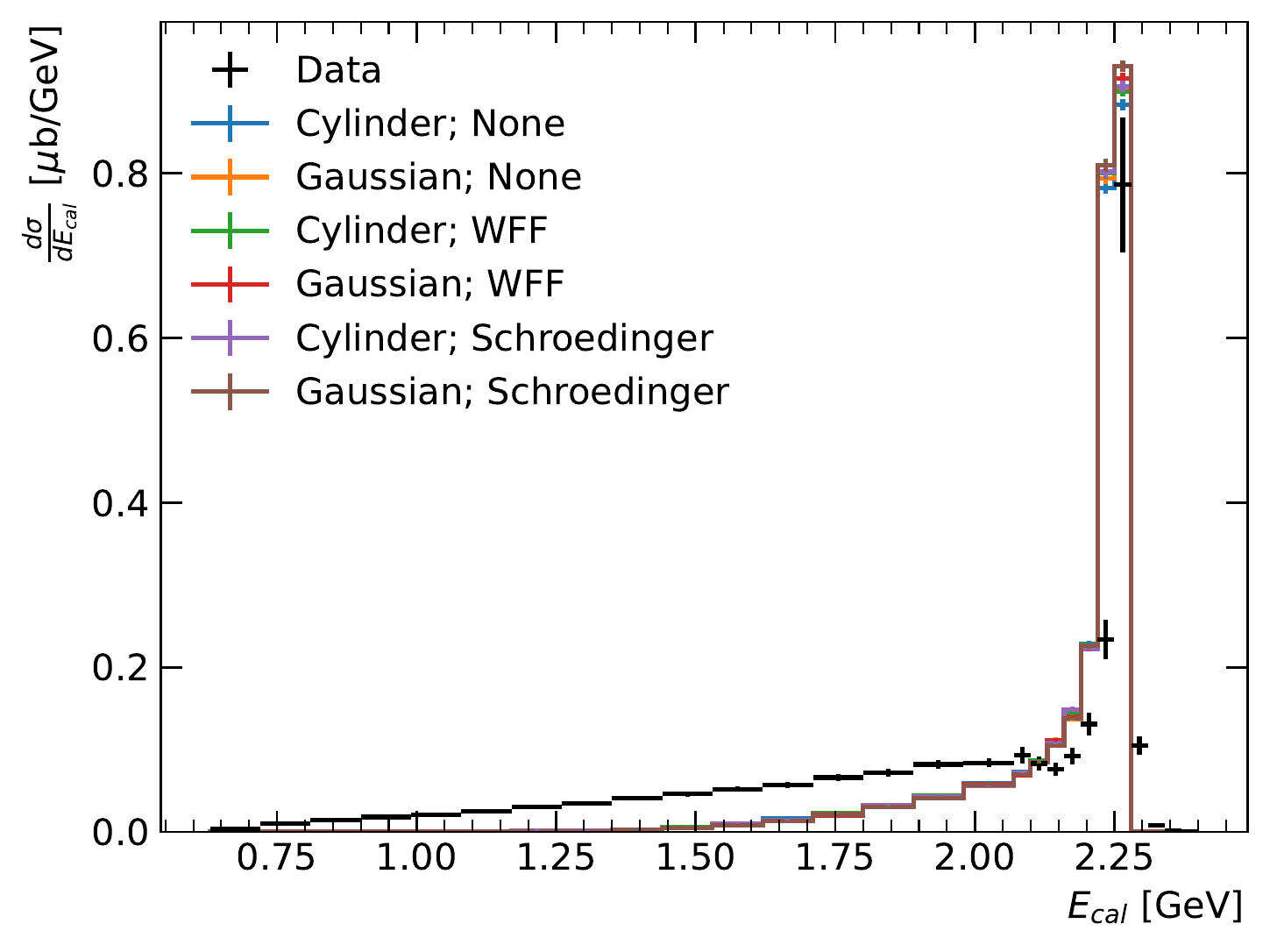}
    \includegraphics[width=0.4\textwidth]{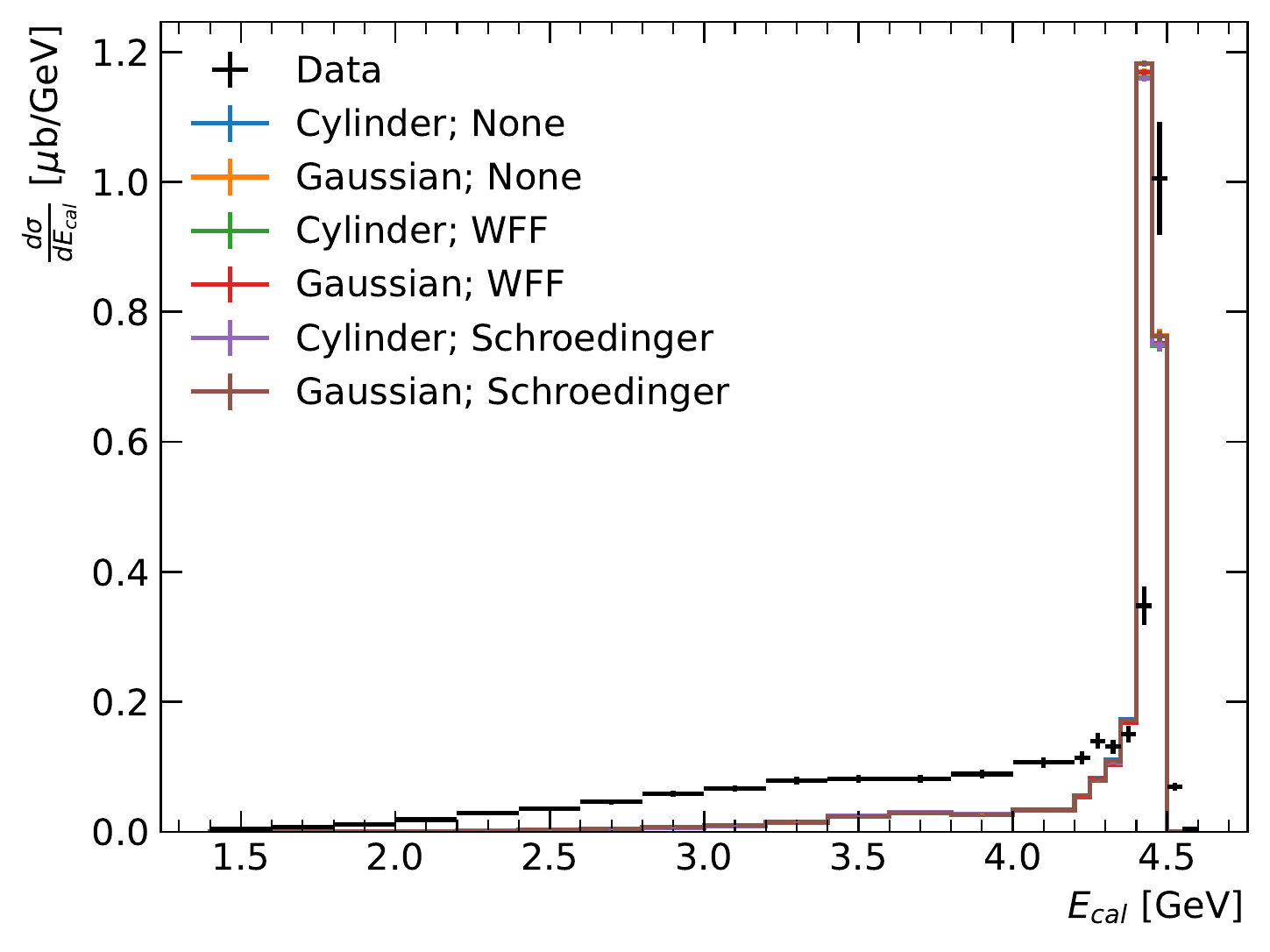}
    \caption{Comparison of the calorimetric energy reconstructed for an electron beam of 1159 MeV (top),
             2257 MeV (middle), and 4453 MeV (bottom). Data is taken from Ref.~\cite{CLAS:2021neh}.
             The definition of $E_{cal}$ can be found in Eq.~\eqref{eq:ecal}.}
    \label{fig:e4nu:e_cal}
\end{figure}

In liquid argon time projection chamber experiments, such as MicroBooNE and DUNE, the ionization energy is used as a mean to reconstruct the incoming neutrino energy. In this case, the calorimetric energy
is defined as
\begin{equation}\label{eq:ecal}
    E_{\rm cal} = \sum_i (E_i + \epsilon_i),
\end{equation}
where $E_i$ is the energy of the lepton or pions or the kinetic energy of the protons, and $\epsilon$ is the average nucleon
separation energy. In the CLAS data, $E_{\rm cal}$ was calculated for events that contained exactly one proton and
zero pions~\cite{CLAS:2021neh}. The comparison between the different schemes and the data are shown in Fig.~\ref{fig:e4nu:e_cal},
with beam energies of 1159 MeV in the top panel, 2257 MeV in the middle panel, and 4453 MeV in the bottom panel. 
Since neutrons do not contribute to the calorimetric energy, we expect this observable to be sensitive to the modeling of the intranuclear cascade.
The peak of these distributions correspond to the beam energy and is dominated by the quasielastic contribution.
The tail towards lower energies is due to the intranuclear cascade, a result of the proton interacting with other nucleons as it escapes the nucleus, as well as non-quasielastic interactions, which are not currently implemented in \achilles.
Around the peak, the largest difference in peak height due to distinct implementations of the INC is about 7\%.
To be conservative, we will quote the INC theory uncertainty as the largest difference among all INC implementations.

\begin{figure}[t]
    \includegraphics[width=0.4\textwidth]{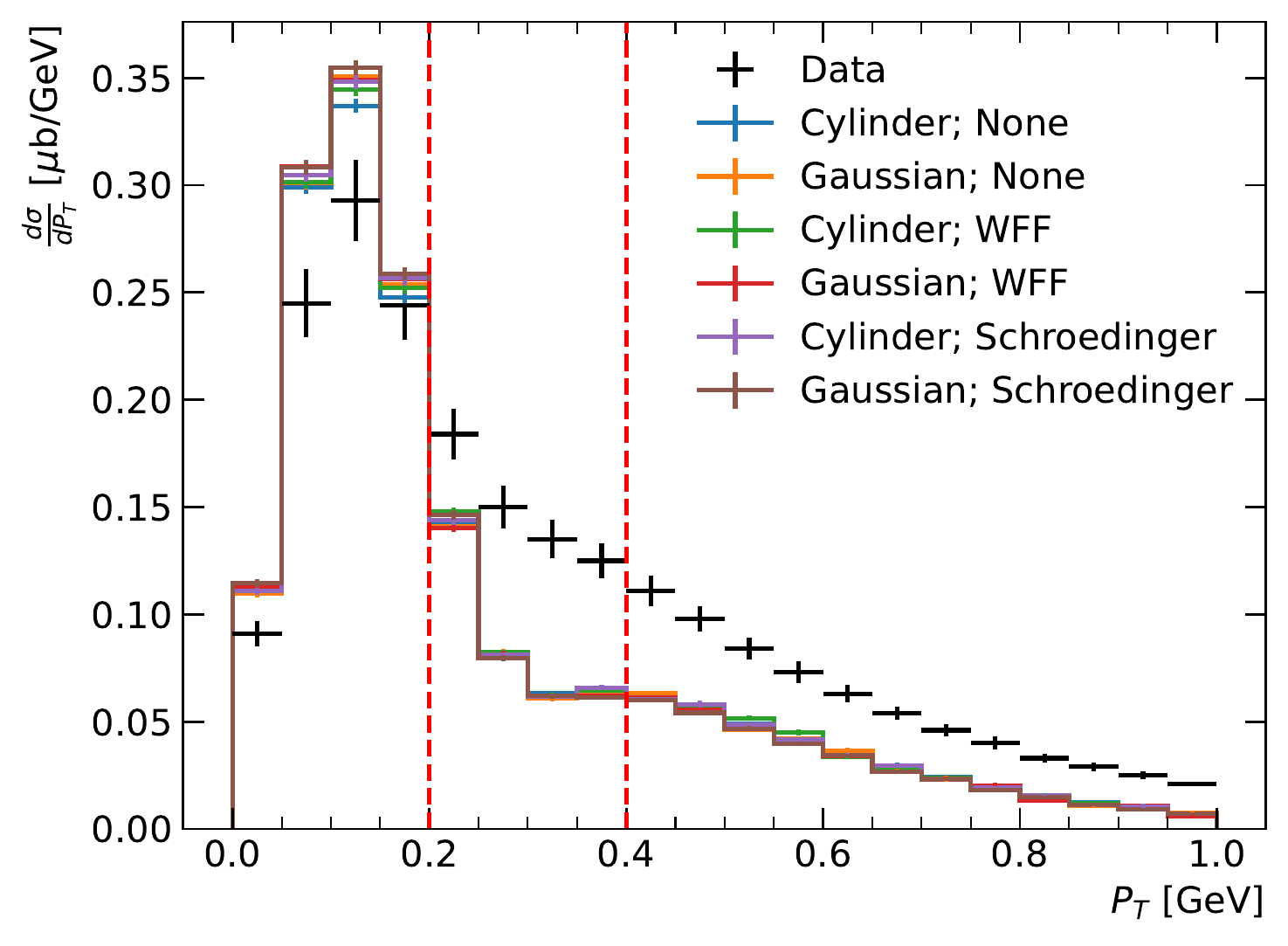}
    \caption{Comparison of the perpendicular momentum for an electron beam of 2257 MeV.
             Data is taken from Ref.~\cite{CLAS:2021neh}.
             The definition of $p^{T}$ can be found in Eq.~\ref{eq:pt}.
             The vertical dashed red lines denote the location of the cuts on $p^T$ at
             200 MeV and 400 MeV.}
    \label{fig:e4nu:pt}
\end{figure}

To further study the accuracy of event simulation, the CLAS and e4v collaborations studied three different transverse momentum
related observables. The first observable is the transverse momentum defined as 
\begin{equation}\label{eq:pt}
    \mathbf{p}^T = \mathbf{p}^T_{e} + \mathbf{p}^T_{p},
\end{equation}
where $\mathbf{p}^T_{e,p}$ is the transverse vector momentum with respect to the beam axis for the electron and proton, respectively.
Note that a lepton scattering off protons at rest would only lead to $p^T\equiv |\mathbf{p}^T|=0$.
Fermi motion of nucleons in the nucleus will lead to a distribution of $p^T$ around 100-200~MeV, while the intranuclear cascade can introduce a long tail towards large values of $p^T$.
This observable is compared in Fig.~\ref{fig:e4nu:pt} to the e4v data. 
The cascade tends to broaden the spectrum 
in the quasielastic region, increasing the maximum value observed. 
The spread due to different INC implementations is about 6\%.
The $p^T>0.2$~GeV region has significant contributions from non-quasielastic processes.

\begin{figure}[t]
    \includegraphics[width=0.4\textwidth]{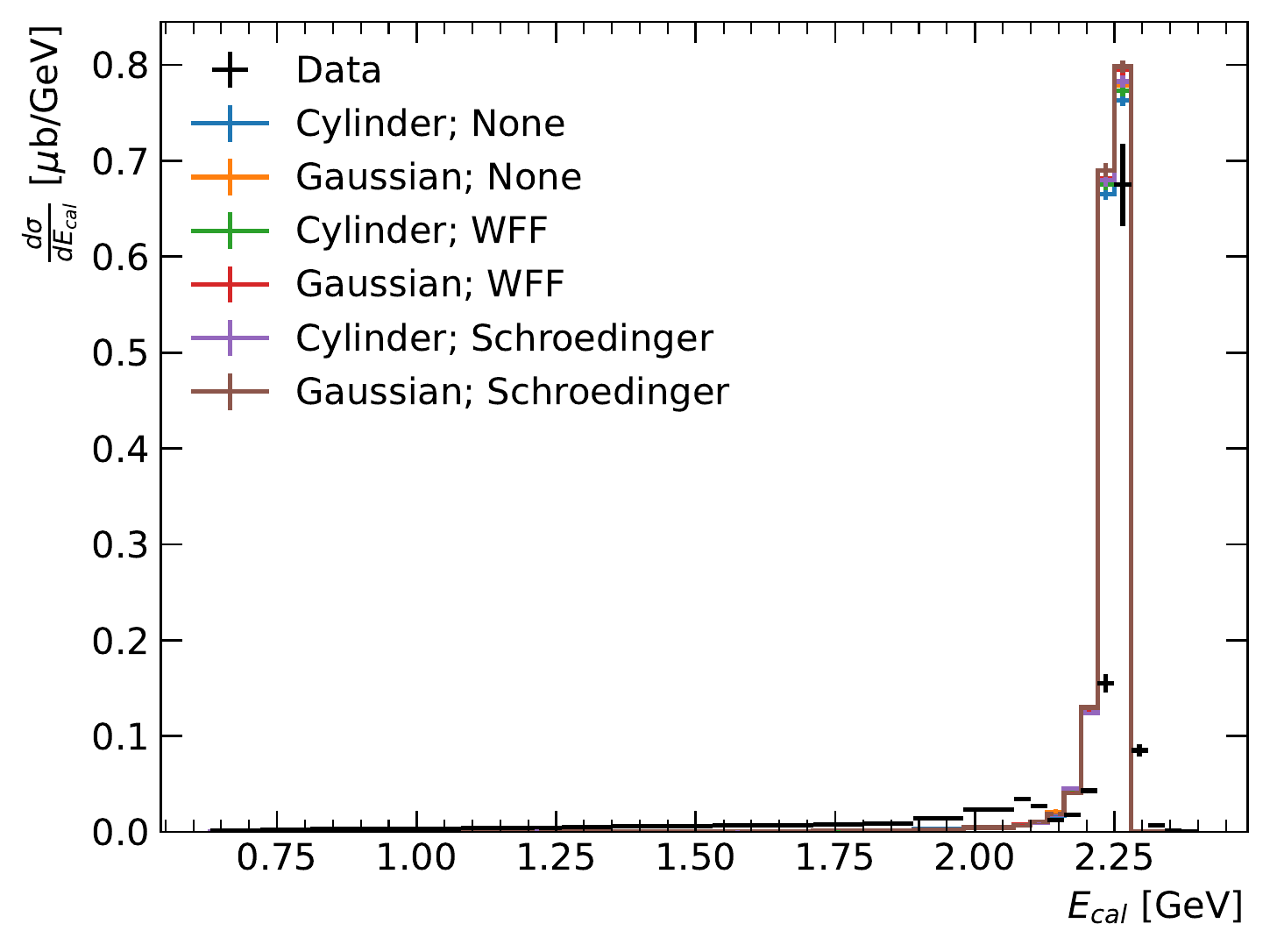}
    \includegraphics[width=0.4\textwidth]{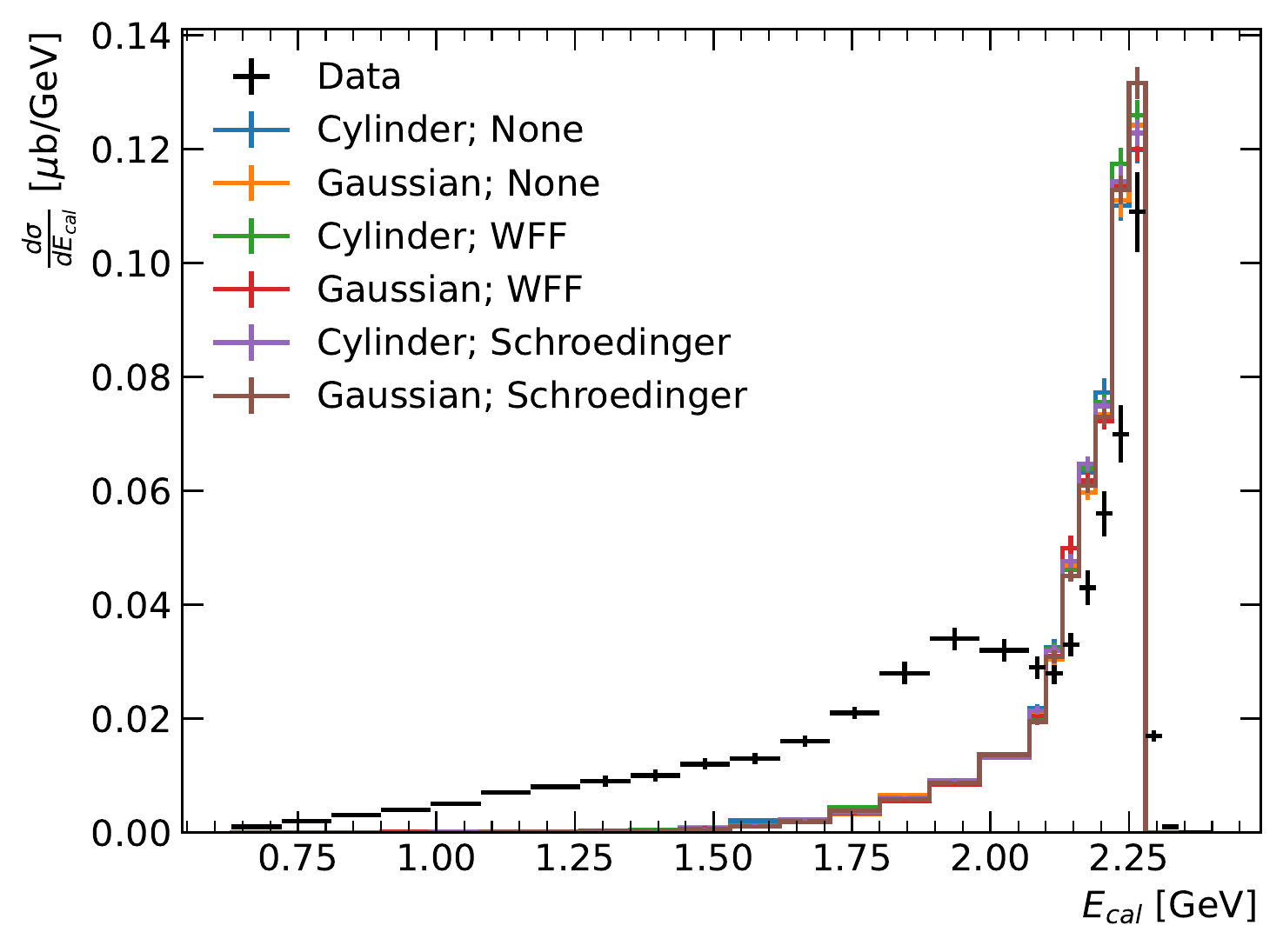}
    \includegraphics[width=0.4\textwidth]{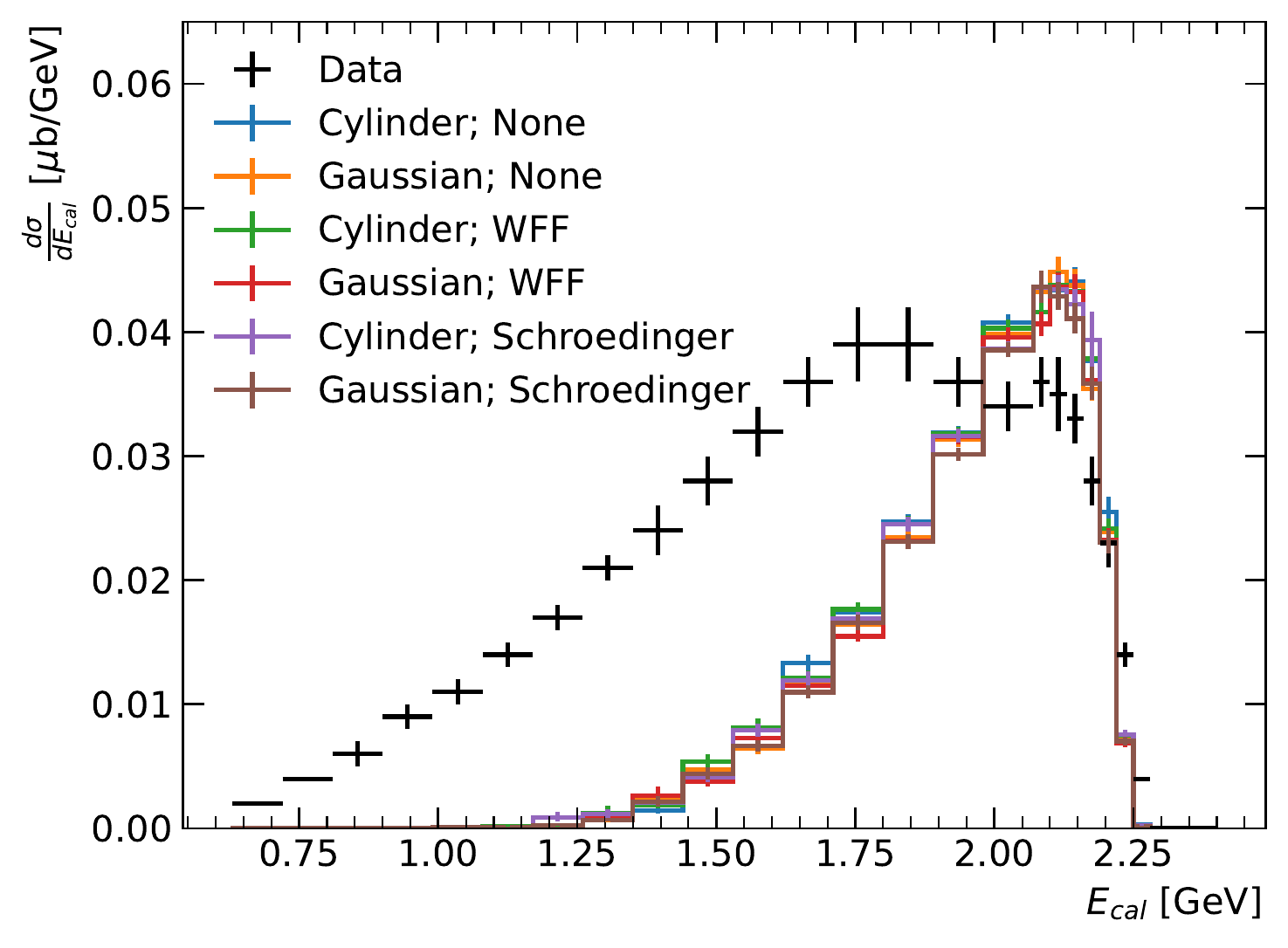}
    \caption{Comparison of the calorimetric energy reconstructed for an electron beam of 2257 MeV, with a cut on 
             the perpendicular momentum of $p^T < 200$ MeV (top), 200 MeV $ < p^T < 400$ MeV (middle),
             and $p^T > 400$ MeV (bottom). Data is taken from Ref.~\cite{CLAS:2021neh}.}
    \label{fig:e4nu:pt_ecal}
\end{figure}

To isolate contributions from different nuclear processes, a cut is applied in the $p^T$ variable before constructing the $E_{cal}$ distributions.
The results are shown in Fig.~\ref{fig:e4nu:pt_ecal} for a cut of $p^T < 200\MeV$ (top panel), $200 \MeV < p^T < 400\MeV$ (middle panel), and $p^T > 400\MeV$ (bottom panel). Again, intranuclear cascades affect the low $E_{cal}$ tail significantly, together with non-quasielastic interactions. 
This is most evident in the $p^T > 400$~MeV plot, 
in which both effects are expected to have large impact.
We find a theory uncertainty associated to the INC implementation of  5\% to 6\% near the peak of all distributions.

\begin{figure}[t]
    \includegraphics[width=0.4\textwidth]{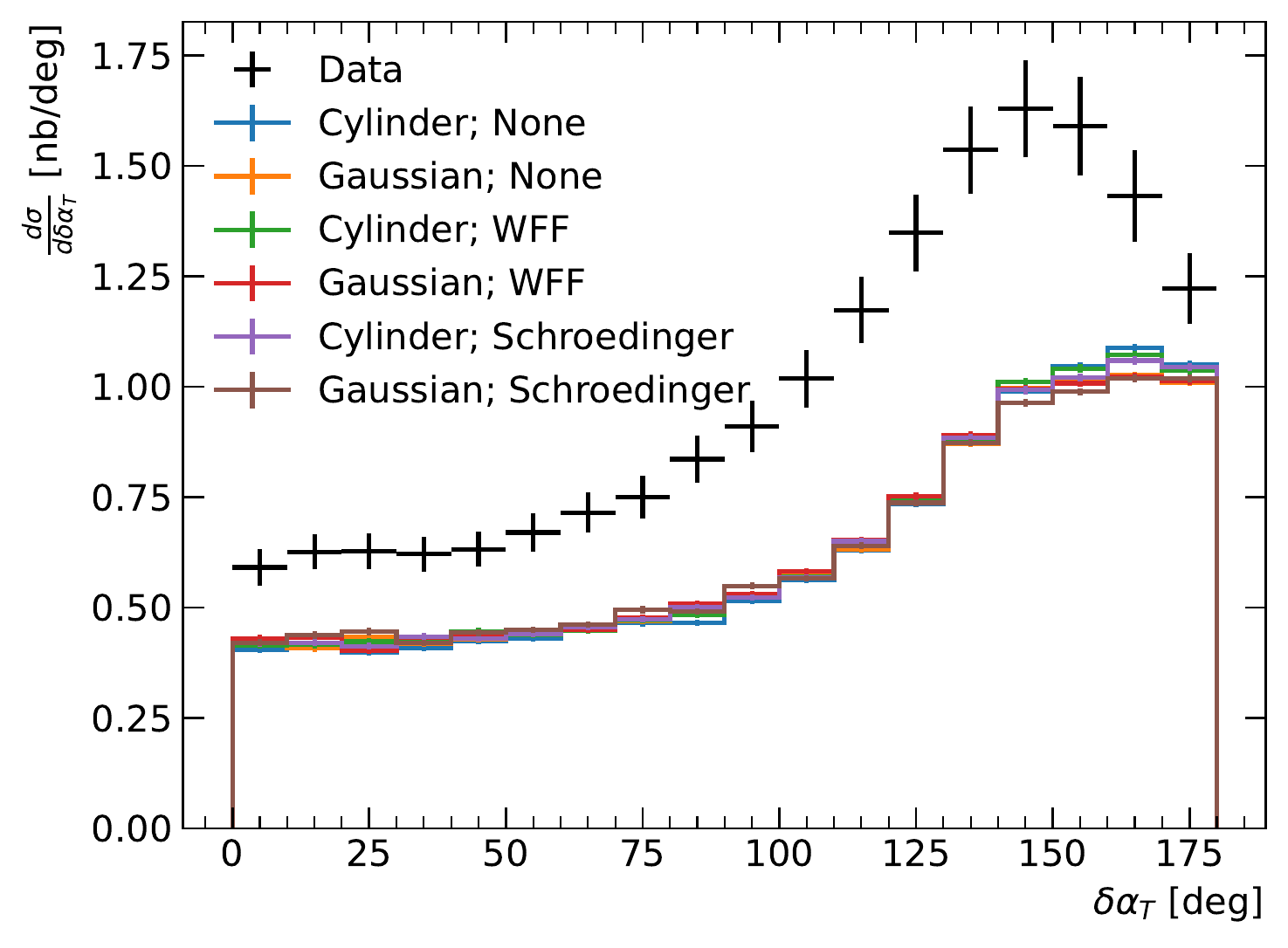}
    \includegraphics[width=0.4\textwidth]{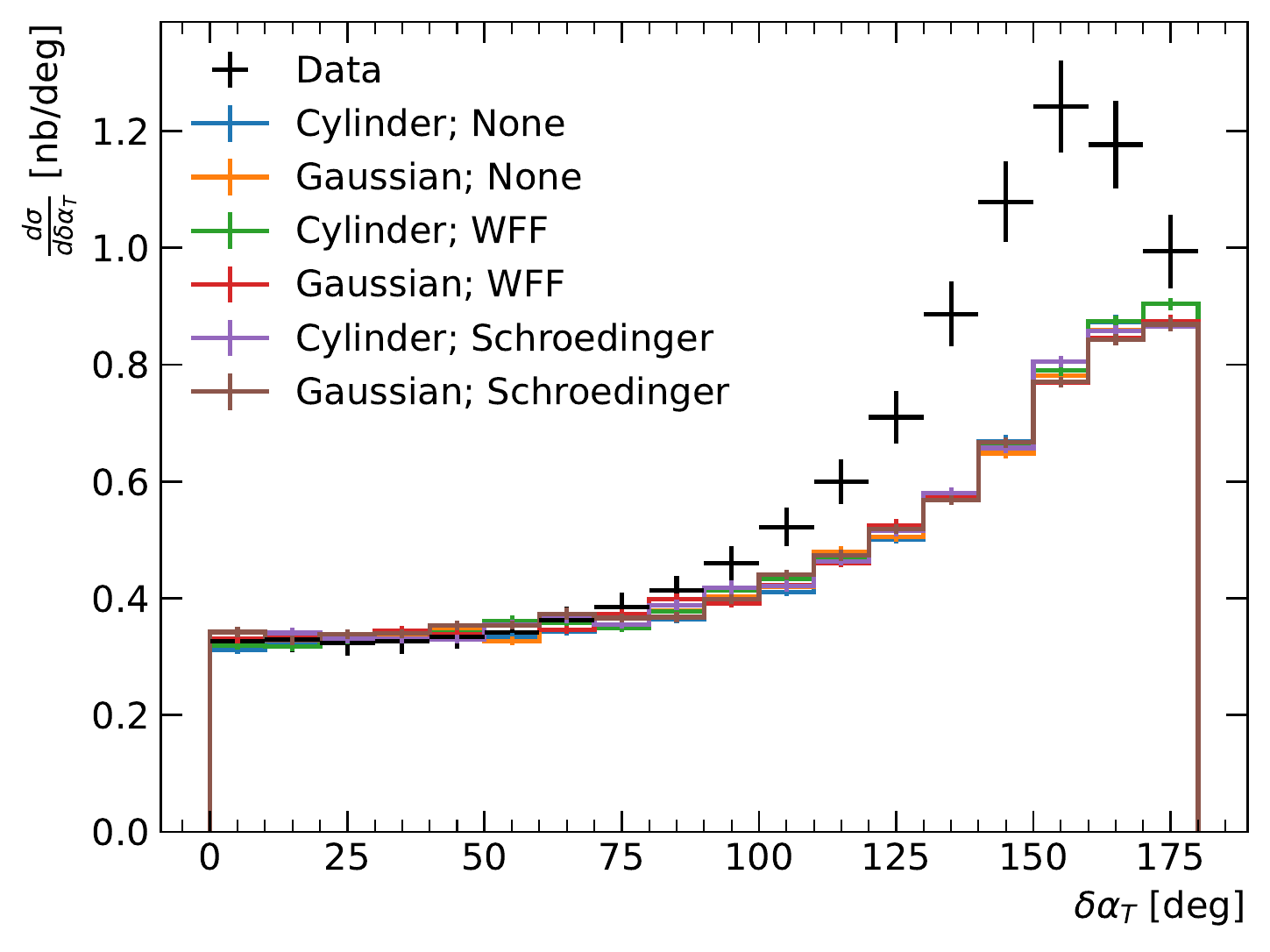}
    \includegraphics[width=0.4\textwidth]{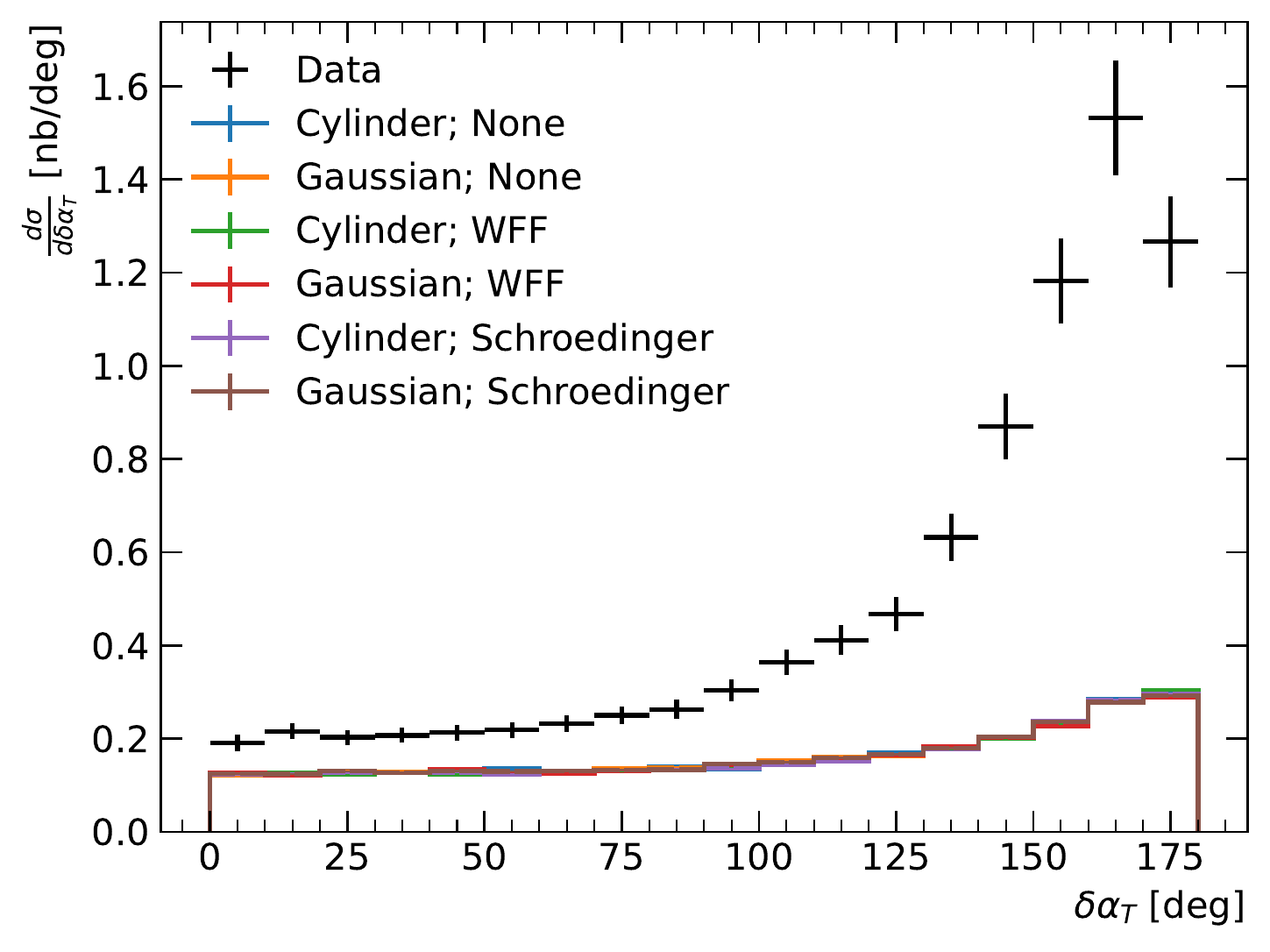}
    \caption{Comparison of $\delta\alpha_T$ for an electron beam of 1159 MeV (top), 2257 MeV (middle),
             4453 MeV (bottom). Data is taken from Ref.~\cite{CLAS:2021neh}.
             The definition of $\delta\alpha_T$ can be found in Eq.~\ref{eq:delta_alpha}.}
    \label{fig:e4nu:alpha}
\end{figure}

The other two observables we use to validate \achilles~are\footnote{Note there are missing minus signs in Eqs. (7) and (8) of Ref.~\cite{CLAS:2021neh}}
\begin{align}
    \delta\alpha_T &= \arccos \frac{-\mathbf{p}^T_{e} \cdot \mathbf{p}^T}{p^T_e p^T}, \label{eq:delta_alpha} \\
    \delta\phi_T &= \arccos \frac{-\mathbf{p}^T_{e} \cdot \mathbf{p}^T_{p}}{p^T_e p^T_p}. \label{eq:delta_phi}
\end{align}
Note that $\mathbf{p}^T=-\mathbf{q}^T$, so $\delta\alpha_T$ is the angle between the overall transverse momentum and the transverse momentum transfer.
Our results for $\delta\alpha_T$ are shown in Fig.~\ref{fig:e4nu:alpha}.
In the limit of no final state interactions, $\mathbf{p}^T$  is simply the initial proton transverse momentum.
Since the initial proton momentum is isotropic,  $\delta\alpha_T$ should also be isotropic in this limit.
The increase in the high-angle region of
the $\delta\alpha_T$ distribution can be attributed to intranuclear cascades and non-quasielastic interactions.
We find an INC theory uncertainty in  $\delta\alpha_T$ of about 10\%.

\begin{figure}[!ht]
    \includegraphics[width=0.4\textwidth]{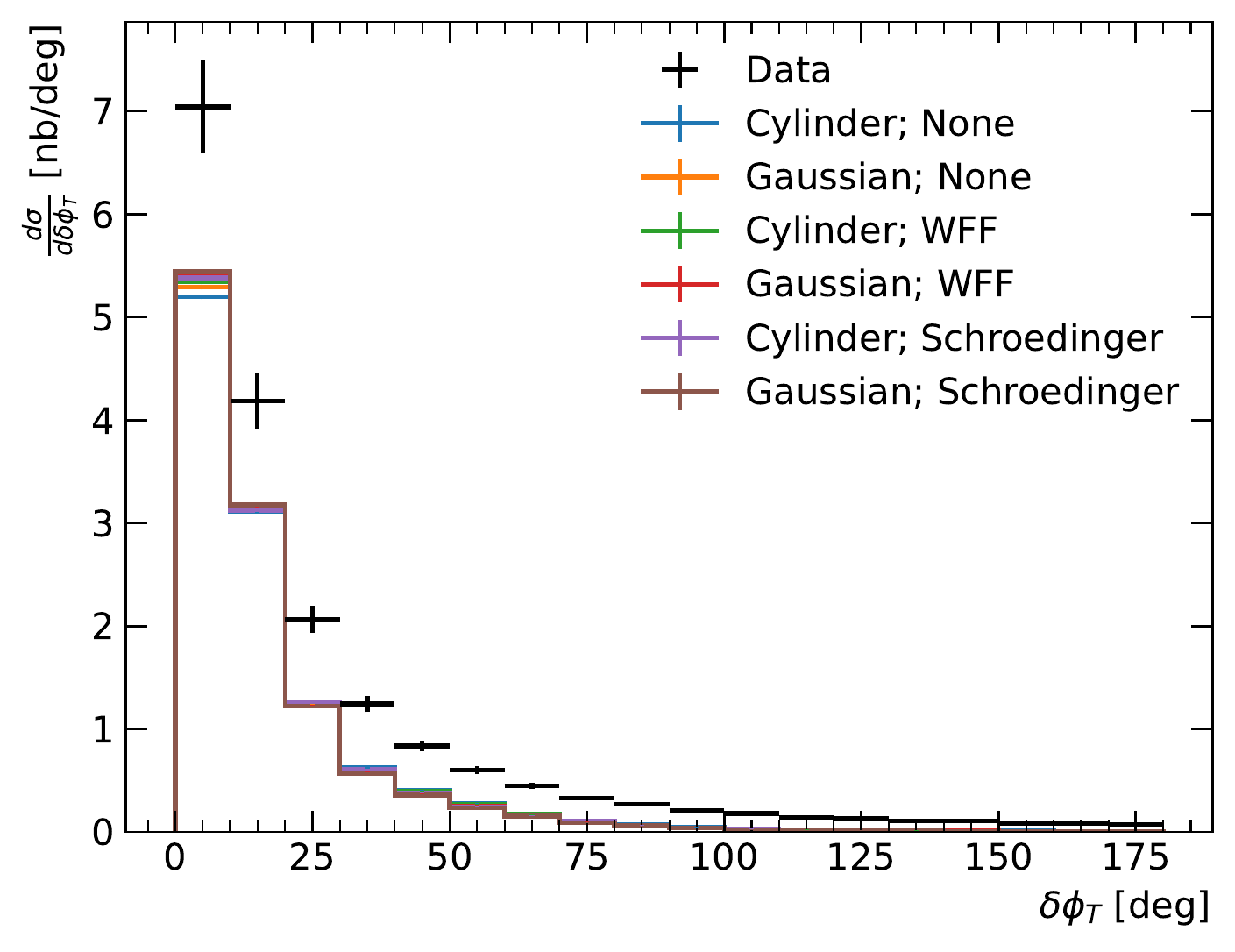}
    \includegraphics[width=0.4\textwidth]{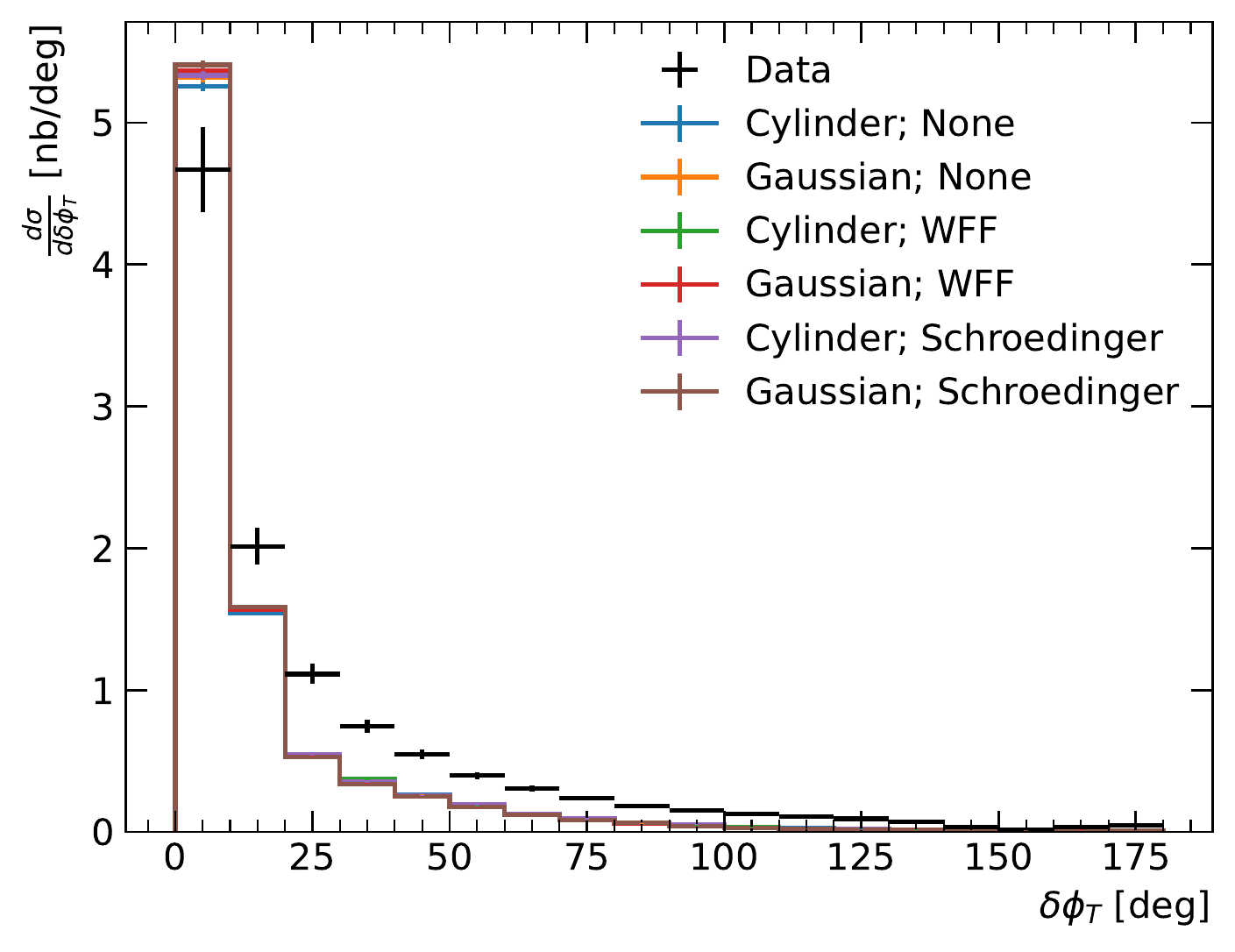}
    \includegraphics[width=0.4\textwidth]{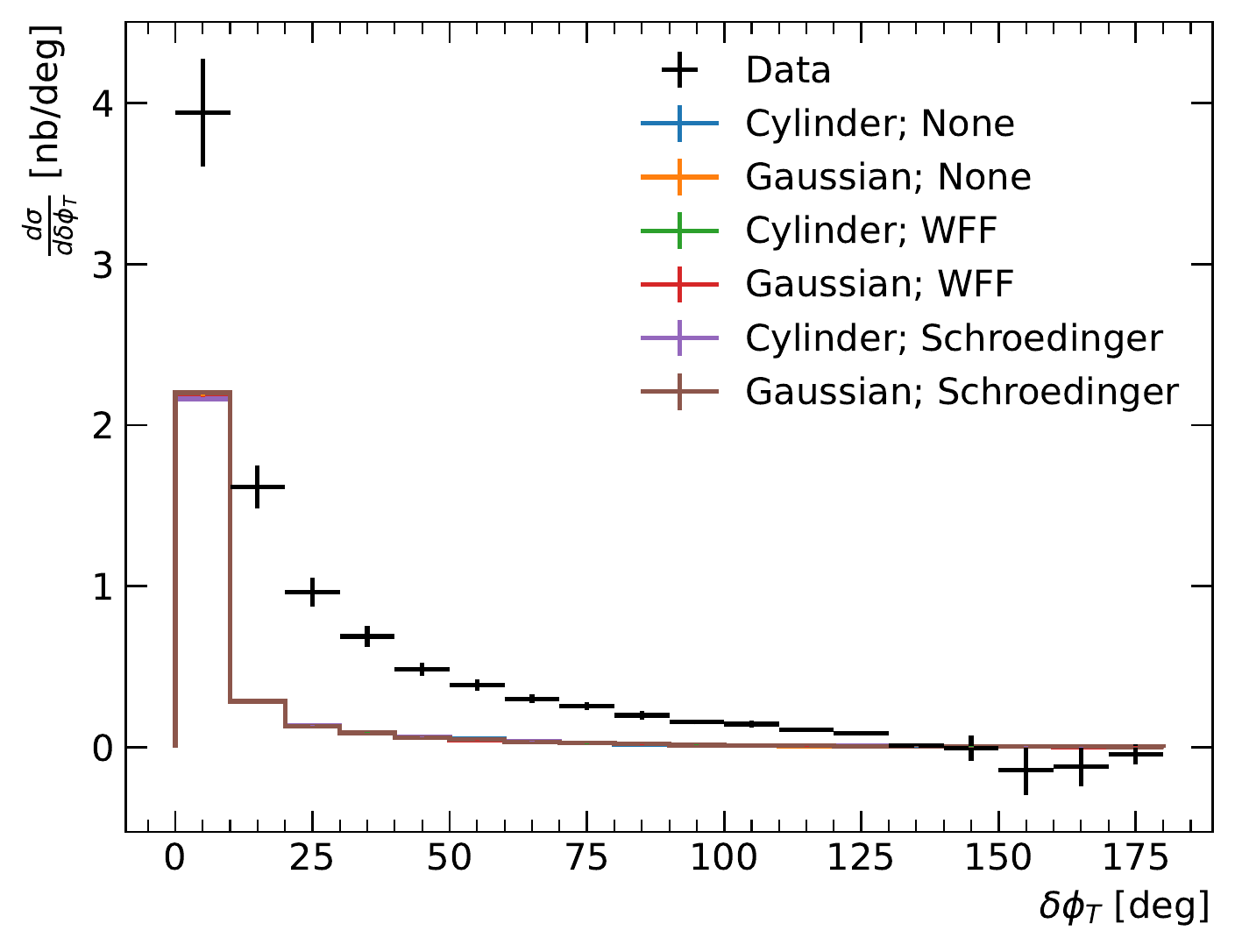}
    \caption{Comparison of $\delta\phi_T$ for an electron beam of 1159 MeV (top), 2257 MeV (middle),
             4453 MeV (bottom). Data is taken from Ref.~\cite{CLAS:2021neh}.
             The definition of $\delta\phi_T$ can be found in Eq.~\ref{eq:delta_phi}.}
    \label{fig:e4nu:phi}
\end{figure}

On the other hand, $\delta\phi_T$
measures the opening angle between the proton and the transverse momentum transfer. 
We present the comparison to data in Fig.~\ref{fig:e4nu:phi}.
In the absence of both final state interactions and initial proton momentum, we have $\mathbf{p}^T_p=-\mathbf{p}^T_e$ and thus $\delta\phi_T$ is a delta function at zero.
In the presence of Fermi momentum, the struck proton has a nonzero momentum, $k_{p}\neq0$, which smears the $\delta\phi_T$ distribution around zero by $\mathcal{O}(k^T_{p}/p^T_e)$.
Final state
interactions help to smear out the distribution to larger opening angles, partially explaining the high $\delta\phi_T$ tails in Fig.~\ref{fig:e4nu:phi}.
The INC uncertainty is found to be about 5\%.

Finally, notice that observables for which final-state interactions play an important role offer the greatest sensitivity to the implementation of the INC model.
This sensitivity is visible in the spread in the color histograms and is particularly evident,
for example, in the $E_{cal}$ distribution for higher $p_T$ (see the bottom panel of Fig.~\ref{fig:e4nu:pt_ecal}) and in the high angle region of the $\alpha_T$ distribution (see top and middle panels in Fig.~\ref{fig:e4nu:alpha}).

\section{Other observables}\label{sec:new_observables}

In this section we propose additional key observables that could be measured in current and future electron-nucleus scattering experiments, such as CLAS12~\cite{Burkert:2020akg} or LDMX~\cite{LDMX:2018cma}.
The goal is to encourage the experimental collaborations to present such observables that will ultimately serve to validate lepton-nucleus interaction models.

Let us start with an exclusive differential observable that is highly sensitive to final state interactions: the \emph{proton multiplicity energy spectrum}.
As we currently do not have pions propagating in our intranuclear cascade modeling in \achilles, we focus on $n$p0$\pi$ events.
Taking the 2.257~GeV electron beam as an example, for every event, we count the number of protons that pass experimental cuts (see Sec.~\ref{sec:QE_exclusive}).
Then we take all leading-energy protons in events with at least one proton and build their energy spectrum.
We repeat the procedure for all second- and third-leading protons, in events with at least two or three protons, respectively.

\begin{figure}[!ht]
        \includegraphics[width=0.45\textwidth]{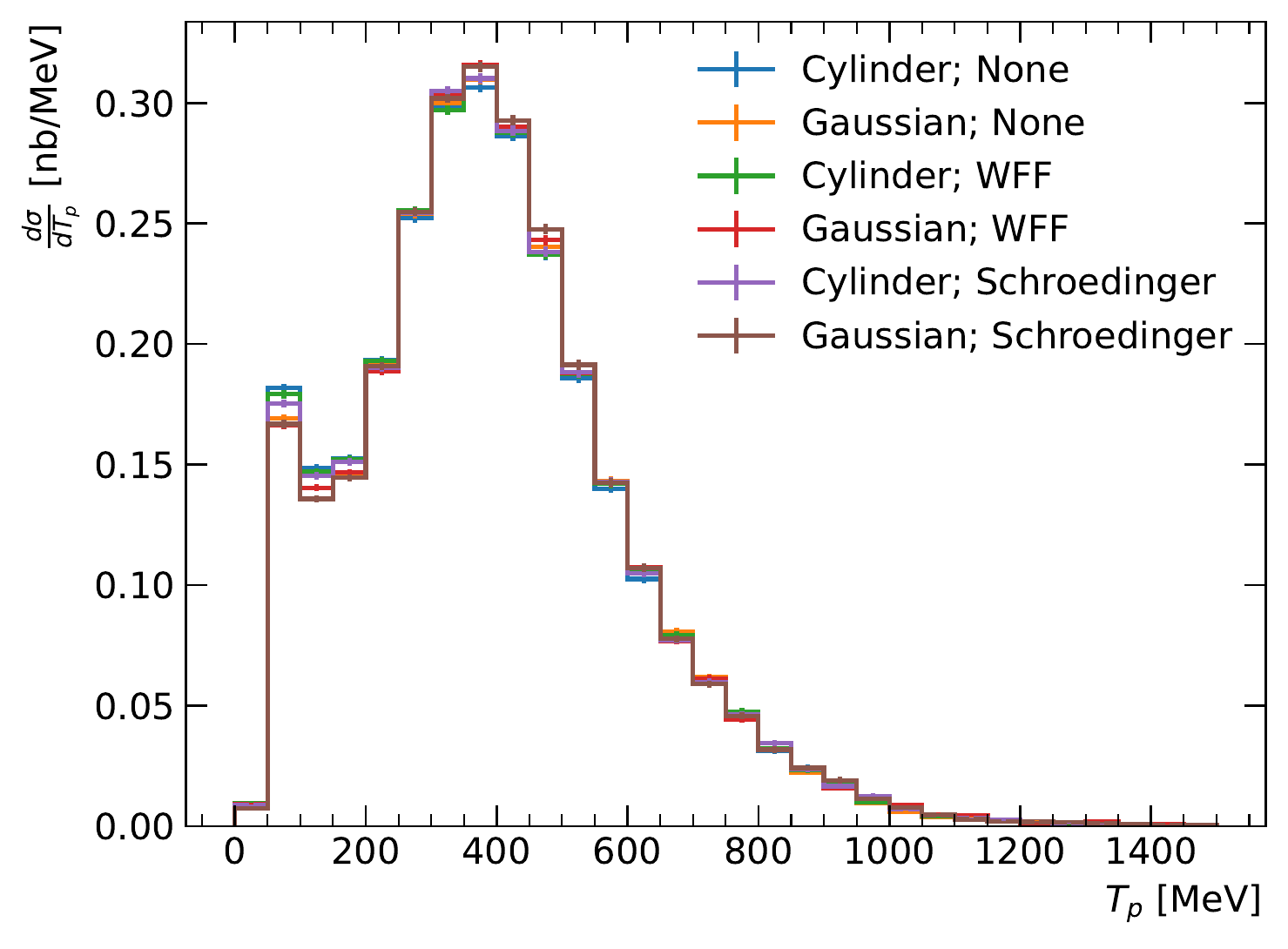}
        \includegraphics[width=0.45\textwidth]{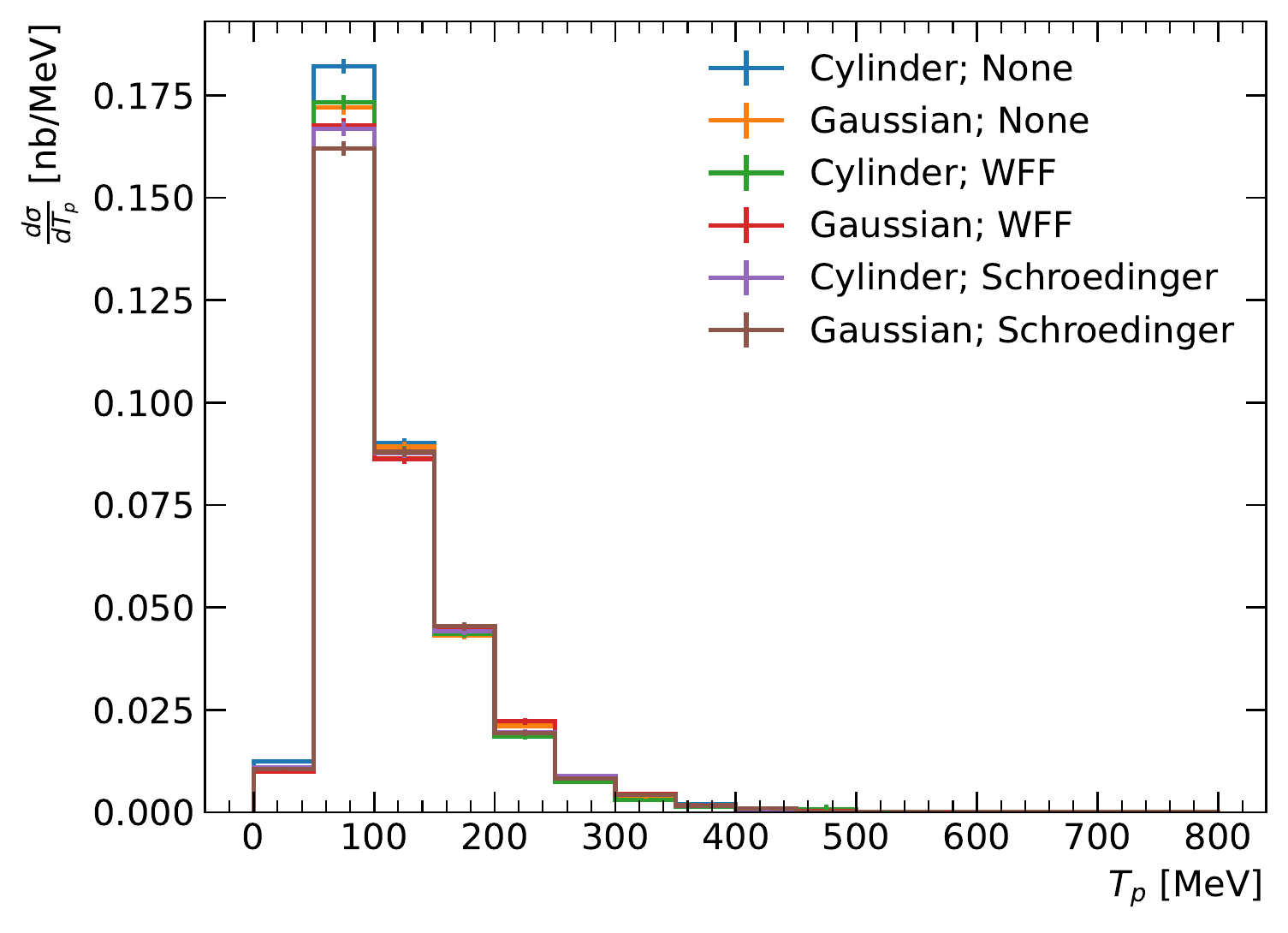}
        \includegraphics[width=0.45\textwidth]{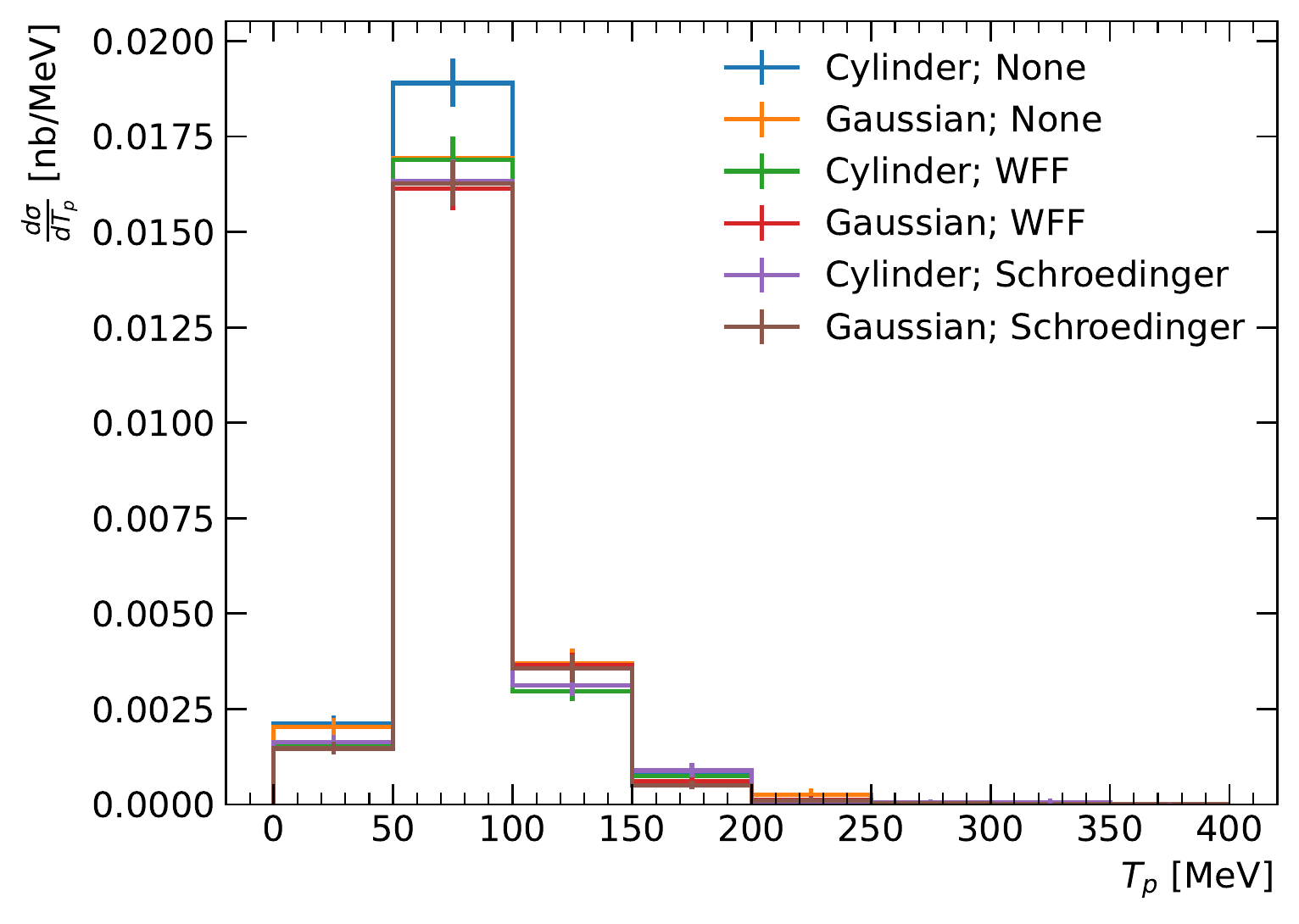}
        \caption{Energy spectra of the $n$-th most energetic proton, from the most (upper panel) to the third most energetic protons (lower panel) for an electron beam with an energy of 2.257 GeV.}
        \label{fig:proton_energy}
\end{figure}

The results of this procedure are the proton energy spectra shown in Fig.~\ref{fig:proton_energy}, from the leading proton in the upper panel to the third leading proton in the lower panel.
We would expect this distribution to be highly sensitive to the intranuclear cascade model.
INCs may raise the proton multiplicity, contributing to the spectra of second and third protons, and tend to distribute the energy among all outgoing protons, shifting the leading proton spectrum towards lower energies.
This is indeed observed when comparing the INC uncertainties in peak regions in the three panels of Fig.~\ref{fig:proton_energy}, which are approximately 3\%, 12\%, and 15\%, from top to bottom.
We also expect that other cross section channels will significantly contribute to this observable in a nontrivial way.
For example, while DIS occurs for higher momentum transfer, hadronization followed by final state interactions may lead to a large multiplicity of low energy protons.
The proton multiplicity energy spectra for all interaction channels will be a subject of a future publication.

\begin{figure}[!ht]
        \includegraphics[width=0.45\textwidth]{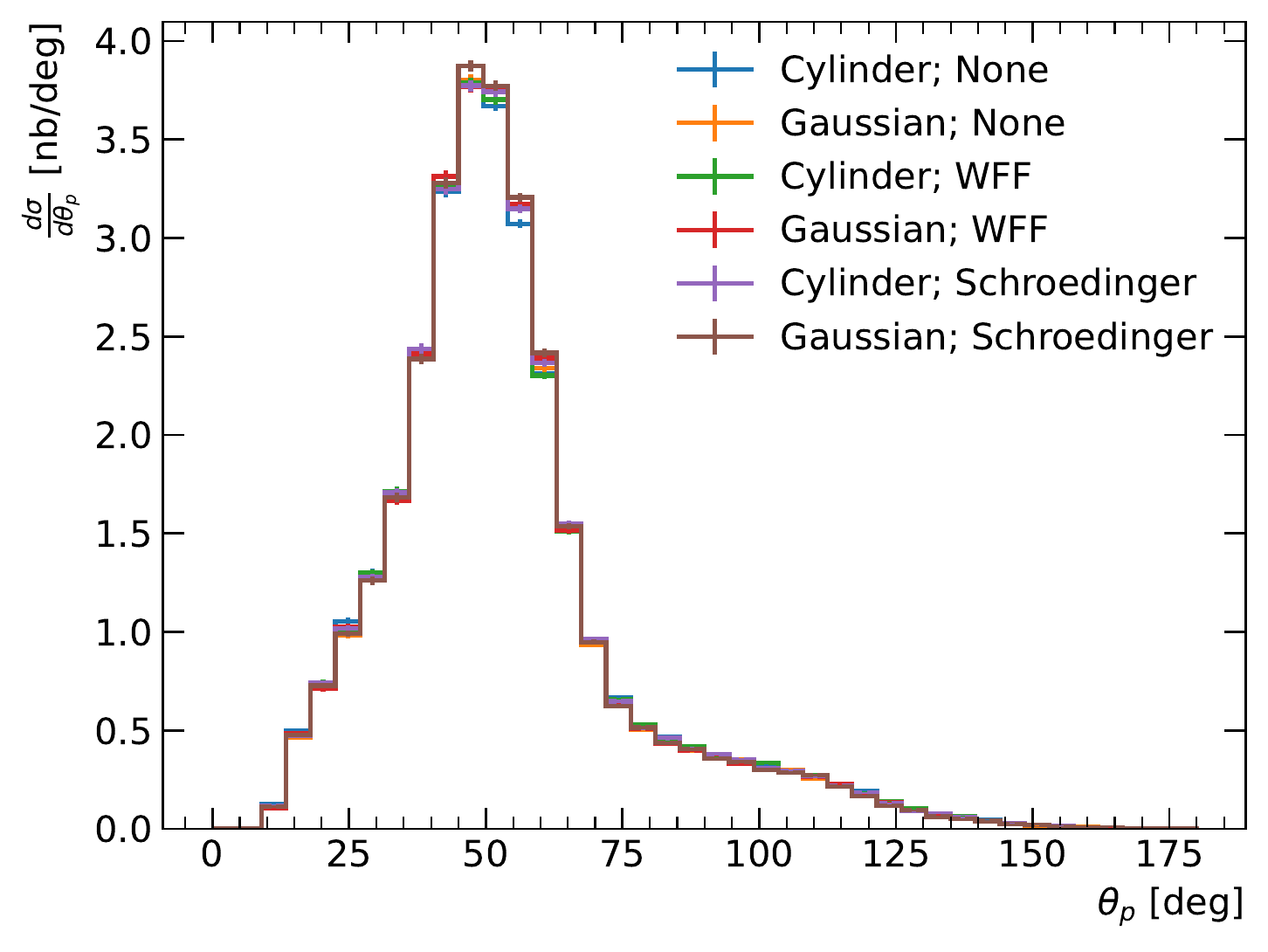}
        \includegraphics[width=0.45\textwidth]{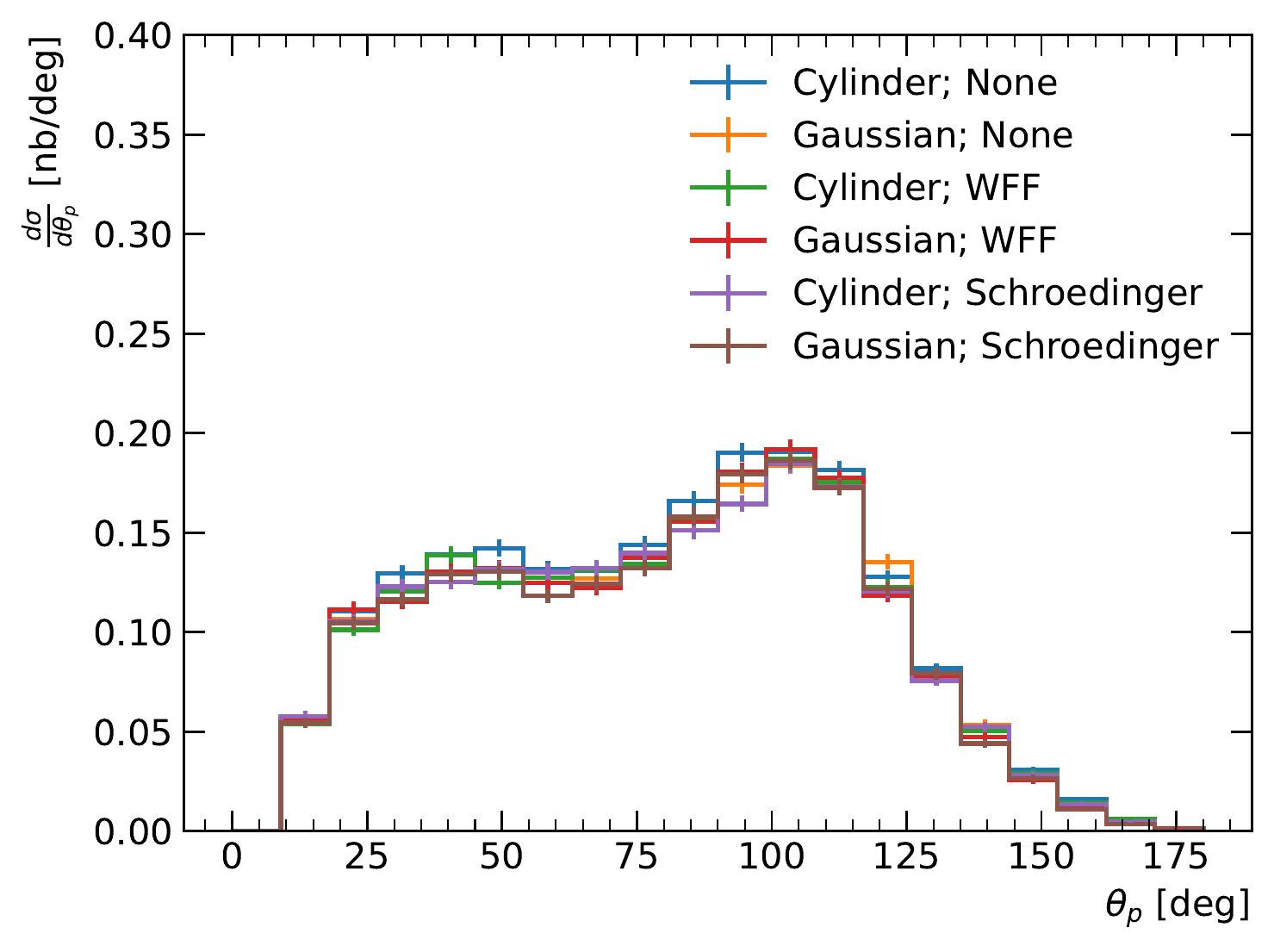}
        \includegraphics[width=0.45\textwidth]{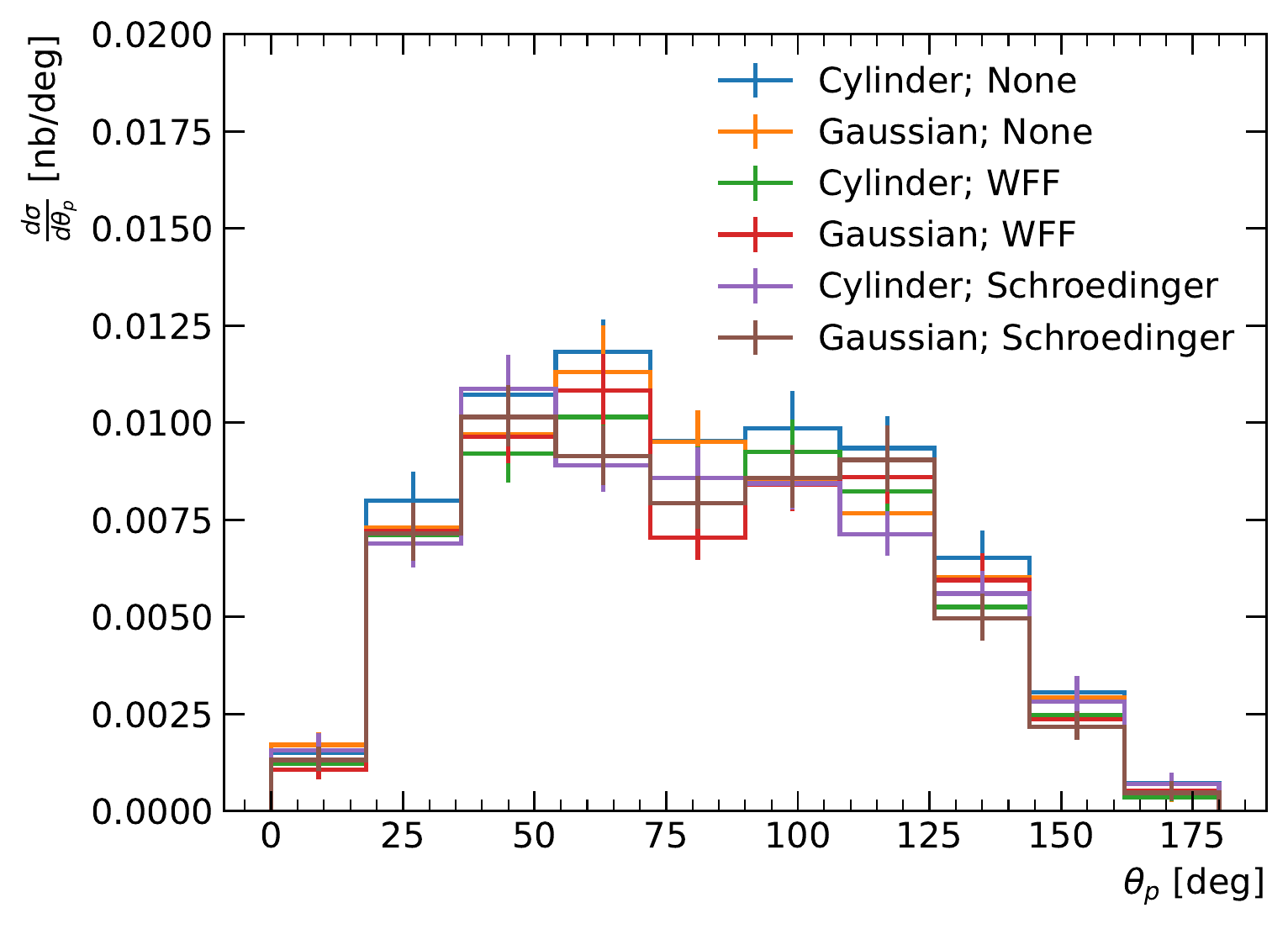}
        \caption{Angular spectra with respect to the beam axis of the $n$-th most energetic proton, from the most (upper panel) to the third most energetic protons (lower panel) for an electron beam with an energy of 2.257 GeV.}
        \label{fig:proton_angle}
\end{figure}

In the same vein, we also propose the \emph{proton multiplicity angular spectra}.
We take again the 2.257~GeV electron beam as an example.
We plot the angle of the leading, second and third protons with respect to the beam axis in Fig.~\ref{fig:proton_angle}, from the leading proton in the upper panel to the third leading proton in the lower panel.
Note that we have decided to order the protons according to their energies.
Our main motivation lies on the fact that higher energy protons are more relevant to the reconstruction of neutrino energies, and therefore a correct description of the leading protons is more relevant than the subleading ones.
Again, we expect intranuclear cascade models to play a crucial role here, as well as the other interactions channels, which will be the studied in a future publication.

Another interesting observable is the angle between the sum of the momenta of all visible outgoing particles with respect to the beam axis.
The only particle we take to be invisible here are neutrons.
We apply the 1p0$\pi$ selection cuts from the CLAS/e4v analysis, see Sec.~\ref{sec:QE_exclusive}.
This angle would be zero in the case of an electron scattering on a free proton at rest.
This observable is motivated the physics of atmospheric neutrinos. 
In this sample, the incoming neutrino direction needs to be reconstructed  in order to estimate the neutrino path through the Earth and in the oscillation probabilities.
A measurement of atmospheric neutrinos in the $0.1-1$~GeV scale at the DUNE experiment could provide nontrivial information on the $CP$ violation phase~\cite{Kelly:2019itm}, and could also be used to perform a tomography study of the Earth, contributing to our understanding of the chemical composition of its core~\cite{Kelly:2021jfs}.

\begin{figure}[t]
        \includegraphics[width=0.45\textwidth]{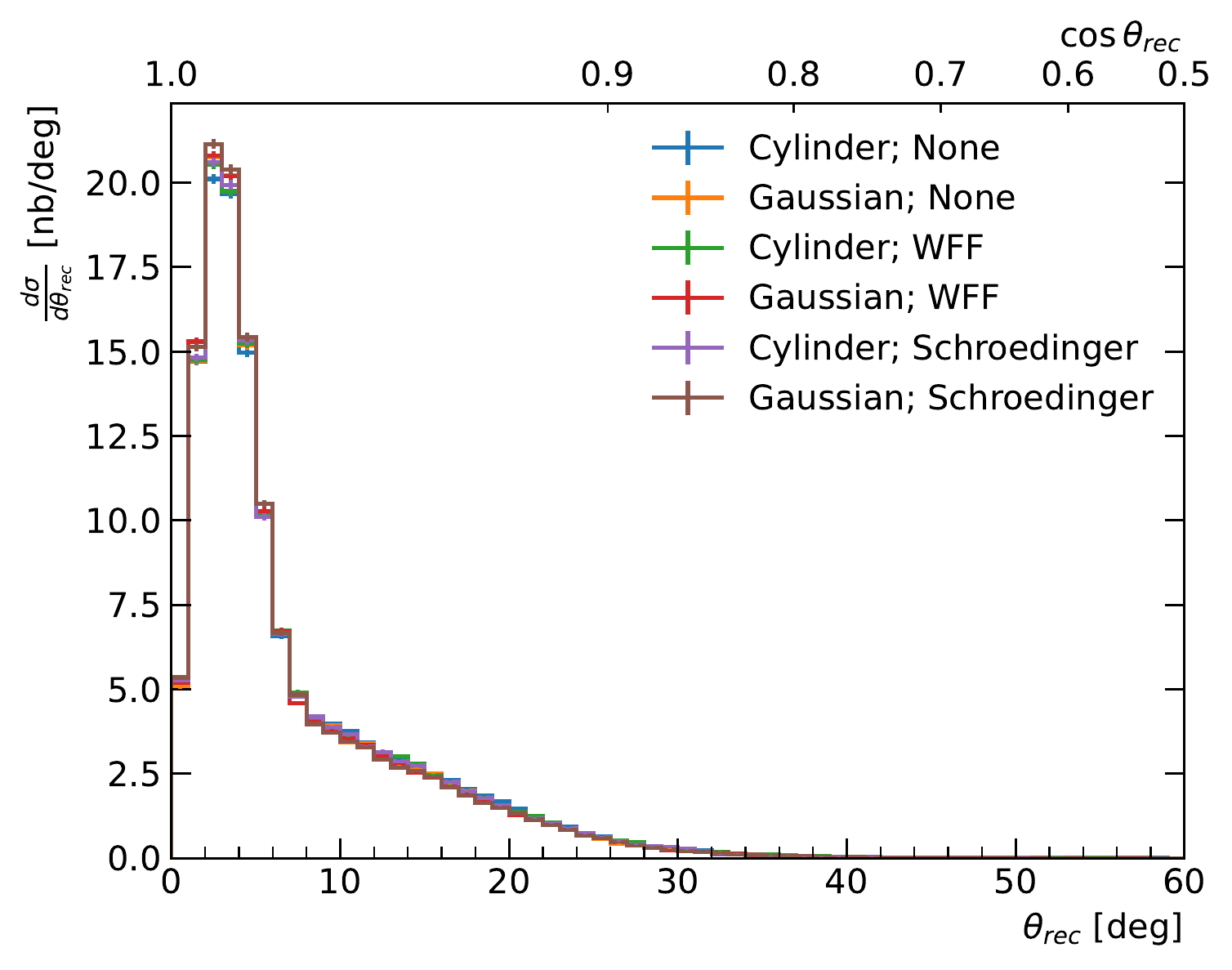}
        \caption{Reconstructed angle of incoming electron with respect to the beam axis for an electron beam with an energy of 2.257 GeV.}
        \label{fig:beam_angle}
\end{figure}

To be more precise, we define the reconstructed beam angle in electron-nucleus scattering as
\begin{equation}\label{eq:theta_incoming}
  \cos\theta_{\rm rec} \equiv \frac{ \hat{\mathbf{k}}_e \cdot \mathbf{p}_{\rm out}}{|\mathbf{p}_{\rm out}|},
\end{equation}
where $\mathbf{p}_{\rm out}$ is the sum of all momenta of visible outgoing particles and $\hat{\mathbf{k}}_e$ is a unit vector in the beam direction.
The reconstructed angle $\theta_{\rm rec}$ can deviate from zero for several reasons: Fermi motion, as it adds momentum to the incoming proton that is not accounted for in Eq.~\eqref{eq:theta_incoming}; intranuclear cascade, as a proton may scatter off a neutron which in turn may be invisible to most detectors of interest; and nuclear potential, which may deflect the outgoing proton.
Our results are found in Fig.~\ref{fig:beam_angle}.
The spread around $\theta_{\rm rec}=0$ can be attributed to Fermi motion and is of order $\Delta\theta_{\rm rec}\sim p_F/k_e$, where $p_F$ is the Fermi momentum.
The tail at large angles is due to final state interactions, as protons may scatter off neutrons and deflect significantly.
Non-quasielastic interactions should further populate the high $\theta_{\rm rec}$ region.

\section{Conclusions}\label{sec:conclusions}
We have presented a newly developed lepton-nucleus event generator, \achilles.
Our generator factorizes the primary interaction vertex from the propagation of hadrons througout the nucleus, allowing for a great deal of modularity, which is one of the pillars of \achilles.
Due to this modularity, \achilles\ can be used for generating either electron-nucleus or neutrino-nucleus scattering events, and the implementation of numerous  scenarios for physics beyond the Standard Model is straightforward.

We have validated quasielastic scattering against high quality, inclusive and exclusive, electron-carbon data, including the recent CLAS/e4v reanalysis of existing data.
We find good agreement between data and simulation.
A complete estimate of the theoretical uncertainty associated with the nuclear model and the current operator adopted in the description of the primary interaction vertex is highly non-trivial and has not been included in this work. 
A promising avenue to quantify model dependence involves testing different nuclear many-body methods, possibly including different nuclear currents, form factors, and Hamiltonians as inputs.
A study along these lines has been carried out in Ref.~\cite{Andreoli:2021cxo} where the inclusive differential cross sections for electron scattering on $^3$He and $^3$H have been evaluated using different many-body approaches based on the same description of nuclear dynamics.
Inputs from LQCD calculations, such as nucleon form factors and elementary nucleon matrix elements, will be incorporated as they become available.

By varying model assumptions of the intranuclear cascade (namely, different nucleon-nucleon interactions models and nuclear potentials), we have estimated one component of the overall theory uncertainty budget in electron-nucleus scattering.
For observables that are sensitive to final-state interactions, the theoretical model dependence associated with different intranuclear cascade models is typically a 5-10\% effect.
Theory uncertainty estimates will be crucial for a precision neutrino physics program, in particular for the DUNE experiment.

We have also proposed novel observables that will allow for further validation of lepton-nucleus scattering models.
Although we have only analyzed electron-carbon scattering data in the quasielastic region, our code is readily extendable to generate neutrino-nucleus scattering events.
Comparison against neutrino scattering data, as well as the inclusion of other primary interaction modes, such as resonant scattering, meson exchange current and deep inelastic scattering, will be subjects of future publications.


\section{Acknowledgments}
We thank Or Hen and Afroditi Papadopoulou for useful discussions. We thank Adi Ashkenazi, Or Hen, Stefan H\"oche, Shirley Li, Kendall Mahn, Afroditi Papadopoulou, Luke Pickering, 
and Larry Weinstein for comments on the manuscript.
Fermilab is operated by the Fermi Research Alliance, LLC under contract No. DE-AC02-07CH11359 with the United States Department of Energy.
The present research is supported by the U.S. Department of Energy, Office of Science, Office of Nuclear Physics, under contracts DE-AC02-06CH11357 (A.L). A.L acknowledges funding from the INFN grant INNN3, and from the European Union’s Horizon 2020 research and
innovation programme under grant agreement No 824093.
This material is based upon work supported by the U.S. Department of Energy, Office of Science, Office of Nuclear Physics under grant Contract Numbers DE-SC0011090 and DE-SC0021006 (W.J.). 
This project has received support from the European Union’s Horizon 2020 research and innovation programme under the Marie Skłodowska-Curie grant agreement No 860881-HIDDeN.

\clearpage
\newpage

\appendix
\section{\texorpdfstring{\achilles}{Achilles} Technical Details}\label{app:achilles}

The calculation of the primary interaction within the \achilles generator is separated into a 
leptonic and a hadronic current as described in~\cite{Isaacson:2021xty}. The use of currents reduces
the bookkeeping required to properly handle the interference between different gauge bosons contributing
in the primary interaction. This is important when dealing with BSM scenarios in which the dominant
contribution to the total cross section may arise from the interference with the Standard Model.
The setup of initial and final state particles, along with the configuration of all other options within the generator is controlled with a set of YAML files. Almost all parameters can be controlled through the YAML input card without the need to recompile the code.

While the core of \achilles is written in modern \texttt{C++} for high performance, a general purpose \texttt{fortran90} wrapper is provided to interface \achilles to available nuclear models.
The wrapper consists of three components. Firstly, the \achilles code provides an interface to the physical constants and other useful common utilities, such as the handling of particle information (particle id, mass, four-momentum, position, status code, etc.).
This helps to ensure consistency of physical constants and particles throughout the calculation. Secondly, the \achilles code expects the nuclear model to define two functions that define the interface between the \texttt{C++} and \texttt{fortran90}. 
The first function handles the initialization of the nuclear model, which gets passed as an argument, a filename and length to be handled by the \texttt{fotran90} code. The second function is expected to perform the calculation of the nuclear current, as discussed in the previous paragraph. 
This function is passed as input information about the four momentum of all the nucleons and the gauge boson for a given event, and expects to be returned the nuclear current. Finally, the nuclear model needs to be registered with the \achilles code to provide a means to enable simulations of the model via the input card.
This wrapper was used to include the original, extensively validated spectral function codes written in
\texttt{fortran90} to be used for the nuclear initial state into the \achilles generator.

The sampling of the phase space is performed with the efficient multi-channel~\cite{Kleiss:1994qy} and recursive phase space~\cite{Byckling:1969sx} discussed in~\cite{Isaacson:2021xty} with importance sampling handled by the VEGAS algorithm~\cite{Lepage:1977sw,Lepage:1980dq}.
This makes the code readily extendable to other reaction mechanisms and higher dimensional phase spaces. Based on the experience of the LHC community, we do not expect any dramatic decrease in computational speed as the multiplicity of the final state increases beyond the extra time involved in evaluating the matrix element and generating the additional momentum. Finally, the major benefit in using these sampling techniques is in the increased unweighting efficiency during the event generation process.

Further details on the input card and API details on the interface will be expanded upon in a manual to be released in the future.

\section{Symplectic Integrator}\label{app:symplectic_int}

As detailed in Ref.~\cite{PhysRevE.94.043303}, in order to develop an explicit symplectic integrator for non-separable Hamilitonians can be achieved by using an augmented Hamiltonian defined as follows:
\begin{equation}
    \overline{H}(q, p, x, y) \equiv H_A(q, y) + H_B(x, p) + \omega H_C(q, p, x, y),
\end{equation}
where $H_A(q, y)$ and $H_B(x, p)$ are copies of the original Hamiltonian, $H_C(q, p, x, y) = |q - x|^2/2 + |p - y|^2/2$ acts as a harmonic oscillator keeping the two solutions close in phase space, and $\omega$ is a tunable parameter to control the strength of the coupling. Explicit flows can then be defined as:
\begin{align}
    \phi_{H_A}^\delta&: \begin{bmatrix} q \\ p \\ x \\ y \end{bmatrix} \to \begin{bmatrix} q \\ p - \delta \partial_{q} H(q, y) \\ x + \delta \partial_{y} H(q, y) \\ y \end{bmatrix}, \\
    \phi_{H_B}^\delta&: \begin{bmatrix} q \\ p \\ x \\ y \end{bmatrix} \to \begin{bmatrix} q + \delta \partial_{p} H(x, p) \\ p \\ x \\ y - \delta \partial_{x} H(x, p) \end{bmatrix}, \\
    \phi_{H_C}^\delta&: \begin{bmatrix} q \\ p \\ x \\ y \end{bmatrix} \to \frac{1}{2}\begin{bmatrix} \begin{pmatrix} q + x \\ p + y \end{pmatrix} + R(\delta) \begin{pmatrix} q - x \\ p - y \end{pmatrix} \\
                                                                                                      \begin{pmatrix} q + x \\ p + y \end{pmatrix} - R(\delta) \begin{pmatrix} q - x \\ p - y \end{pmatrix} \end{bmatrix},
\end{align}
where $\delta$ is the time step for the evolution, and 
\begin{equation}
    R(\delta) \equiv \begin{bmatrix} \cos(2\omega\delta) & \sin(2\omega\delta) \\ -\sin(2\omega\delta) & \cos(2\omega\delta) \end{bmatrix}.
\end{equation}
A second order method can be created from these three Hamiltonians through the use of the symmetric Strang splitting method~\cite{doi:10.1137/0705041}. This leads to a single evolution step over a time step $\delta$ as:
\begin{equation} 
    \phi_2^{\delta} = \phi_{H_A}^{\delta/2} \circ \phi_{H_B}^{\delta/2} \circ \phi_{\omega H_C}^{\delta} \circ \phi_{H_B}^{\delta/2} \circ \phi_{H_A}^{\delta/2}. \label{eq:evolution}
\end{equation}
A $l$th order integration method can be obtained through the triple jump method~\cite{YOSHIDA1990262,mclachlan_quispel_2002}:
\begin{align} 
    \phi_{l}^{\delta} &= \phi_{l-2}^{\gamma_l\delta} \circ \phi_{l-2}^{(l-2)\gamma_l\delta} \circ \phi_{l-2}^{\gamma\delta},\\
    \text{where}\ \gamma_l &= \frac{1}{2-2^{1/(l+1)}}, \nonumber
\end{align}
which will also be symplectic if $\phi_{l-2}$ is symplectic. In this work, we tune the values of $\omega$ and $\delta$ such that the results are stable as a second order integrator.

\begin{center}
\begin{figure*}[ht!]
    \includegraphics[width=\textwidth]{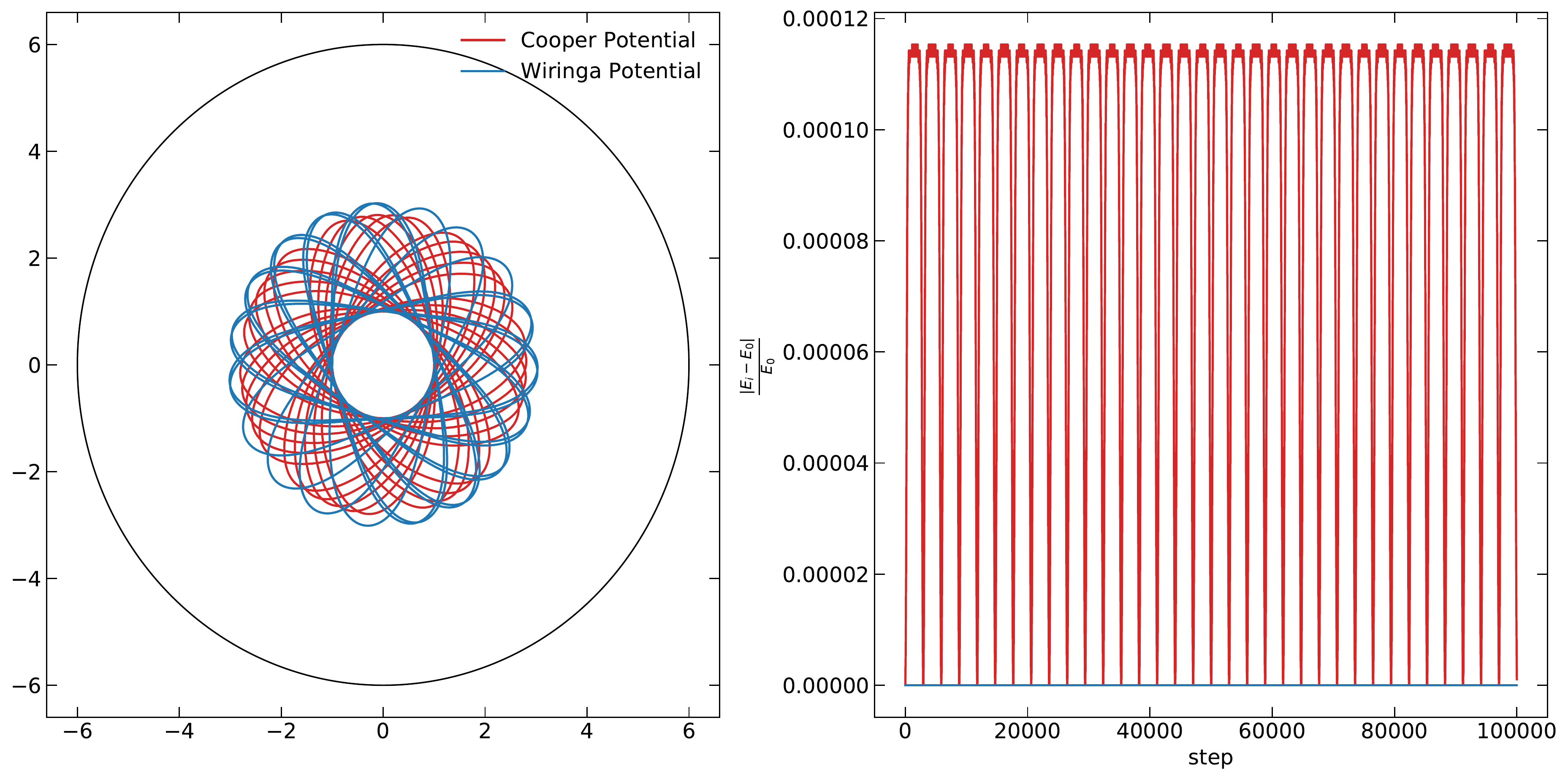}
    \caption{Demonstration of the stability of the symplectic integrator. The left panel shows the trajectory
             of a nucleon with a momentum of 250~MeV perpendicular to the radius starting at a radius of $r=1$~fm in the Wiringa potential (blue) and the Cooper potential (red)
             over 100,000 time steps. The
             right panel shows the deviation from the starting energy as a function of the time step.}
    \label{fig:symplectic}
\end{figure*}
\end{center}

Figure~\ref{fig:symplectic} demonstrates the stability of the symplectic integrator for a nucleon with a momentum of 250~MeV perpendicular to the radius starting at a radius of $r=1$~fm in the non-relativisitc Wiringa potential (blue) and the relativistic Cooper potential (red). The simulation is run for 100,000 time steps, and the maximum energy deviation is of the order of $10^{-4}$. The deviation is periodic in nature, which is a common feature for symplectic integrators.

\bibliography{biblio}

\end{document}